\documentclass[a4paper,12pt]{elsarticle}
\usepackage{lmodern}
\usepackage{color,epsfig,amsmath,amsthm,amssymb,bm,epsf,graphics,graphicx,psfrag,verbatim,subfigure,framed}
\usepackage{mathrsfs}
\usepackage{amsfonts}
\usepackage{verbatim}
\usepackage{listings}
\usepackage{xcolor}
\usepackage{geometry}
\usepackage{url}
\usepackage { algorithm }
\usepackage { algorithmic }
\usepackage{ulem}
\usepackage{mathtools}

\usepackage{soul}

\newcommand{\be}{\begin{equation}}
\newcommand{\ee}{\end{equation}}
\newcommand{\ba}{\begin{eqnarray}}
\newcommand{\ea}{\end{eqnarray}}
\newcommand{\bw}{\begin{widetext}}
\newcommand{\ew}{\end{widetext}}

\newcommand{\rv}{{\bm{r}}}
\newcommand{\xv}{{\bm{x}}}

\journal{Journal of Computational Physics}

\begin{document}

\begin{frontmatter}



\title{A GPU-based Large-scale  Monte Carlo Simulation Method for Systems with Long-range Interactions}


\author{Yihao Liang, Xiangjun Xing}
\address{ Institute of Natural Sciences
 and Department of Physics and Astronomy,
Shanghai Jiao Tong University,
Shanghai, 200240 China
}
\author{Yaohang Li}
\address{Department of Computer Science,
Old Dominion University, Norfolk, VA 23529, United States}
\begin{abstract}
In this work we present an efficient implementation of Canonical Monte Carlo simulation for Coulomb many body systems on graphics processing units (GPU).  Our method takes advantage of the GPU Single Instruction, Multiple Data (SIMD) architectures, and adopts the sequential updating scheme of Metropolis algorithm.  It makes {\it no approximation} in the computation of energy, and reaches a remarkable 440-fold speedup, compared with the serial implementation on CPU. We further use this method to simulate primitive model electrolytes, and measure very precisely all ion-ion pair correlation functions at high concentrations.  From these data, we extract the renormalized Debye length, renormalized valences of constituent ions, and renormalized dielectric constants.  These results demonstrate unequivocally physics beyond the classical Poisson-Boltzmann theory. 

\end{abstract}


\end{frontmatter}


\section{Introduction}
Molecular simulations generally fall into two categories: molecular dynamics (MD) and Monte Carlo (MC). In a MD simulation, one solves the Newtonian equation, from which both dynamical and static properties of studied systems can be extracted.  In a Monte Carlo simulation, one carries out a Markovian stochastic process which converges to the equilibrium Gibbs distribution.  The main advantage of MC is that it can often be accelerated substantially by performing unphysical moves that involve long displacement and/or large number of particles.  Hence MC is usually more efficient than MD for simulation of equilibrium systems.

Long range interactions impose substantial difficulties on numerical simulations, because the computational complexity for one cycle (where every particle moves one step on average) scales as $N^2$, in contrast with $N$ for short range interactions, where $N$ is the size of system being simulated. This severely limits the size of feasible simulations.  There are three classes of methods to speed up the simulation for long range interacting systems: 1) Multipole expansion methods \cite{Tree-code-Binary,Tree-code-Oct,Tree-code-MD-FMM,Tree-code-MC}, where interactions are computed approximately using truncated multipole expansions.  This reduces the computational complexity to $N$ or $N \log N$. 2) Fourier transforms augmented by Ewald-summations, which reduce the complexity to $N^{3/2}$.  It can be further reduced to $N \log N$, by using Fast Fourier Transform. Unfortunately, the latter trick is not applicable for MC. Furthermore, additional artifacts arise due to periodic images.  For a discussion of these artifacts, see reference \cite{Liang-Multi-scale}.  3) Multi-scale reaction-potential methods \cite{Liang-Multi-scale, Xu-Mellin, LBD+:JCP:2009}, whose idea is to simulate only a small portion of the system and use continuum theory to describe the remaining.  These methods have difficulty to scale to system with large number of particles, since they do not use any speedup technique in the computation of energy. 
 
Graphics processing units (GPU) offer a new possibility for speeding up  large scale simulation of long range interacting systems {\it without sacrificing accuracy}.  GPU is a powerful device which can process thousands of threads simultaneously with high memory bandwidth. Compared to CPU, GPU is designed with more transistors that are devoted to data processing rather than data caching and flow control \cite{CUDA-doc}. It is suitable for computation-intensive and data-parallel computations such as graphics rendering - the original purpose of designing GPU. In recent years, the GPU is devoted to more and more general purpose fields, such as data mining, machine learning, finance, scientific computing and molecular simulation.

Many MD simulation methods have already been adapted to GPU in the past years \cite{MD-GPU-Gem3,MD-GPU-Brownian-Dynamics-JCP-2011,MD-GPU-Chamonile-Scheme-2007,MD-GPU-Gravitational-Tree-Code,MD-GPU-Nanjing-2007,MD-GPU-Rigid-Body,MD-GPU-Short-Range-2008,MD-GPU-Papaport}.  There are also many MD softwares and libraries that can be implemented on GPU, including AMBER \cite{MD-GPU-AMBER-1,MD-GPU-AMBER-2,MD-GPU-AMBER-OVERVIEW,MD-GPU-AMBER-SPFP} FENZI \cite{MD-GPU-FENZI,MD-GPU-FENZI-JCC} LAMMPS \cite{MD-GPU-LAMMPS,MD-GPU-LAMMPS-Short-Range} NAMD \cite{MD-GPU-NAMD-STONE-2007,MD-GPU-NAMD-2008}, HALMD \cite{MD-GPU-HALMD}, OpenMM \cite{MD-GPU-OpenMM} HOOMD-blue \cite{MD-GPU-HOOMD-blue-JCP2008,MD-GPU-HOOMD-blue-Website}, GROMACS \cite{GROMACS-Website}, ACEMD\cite{ACEMD-Website}. Most of these packages have demonstrated efficiency in simulating long range interacting systems.

Implementation of Monte Carlo simulation on GPU turns out to be significantly harder.  This is mainly because of the sequential nature of Monte Carlo stochastic dynamics, where particles are moved one by one.
Nonetheless, there have been a few attempts to realize MC simulation of Ising model \cite{GPUIsing2D, GPUIsing2D3D, GPU-MC-Lattice-spin-model} and Hard disk fluid \cite{GPU-MC-Hard-disk} on GPU. 
A. Yaseen and Y. Li used the remapping method to calculate the total energy on GPU for protein systems \cite{Energy-evaluation-GPU}.  A group at Wayne State University \cite{GPU-MC-large-N-1,GPU-MC-large-N-2} realized a GPU code for Gibbs ensemble MC simulation of simple liquids. J. Kim {\it et. al.} developed an implementation with embarrassing parallel on GPU, where each block performs an individual Monte Carlo simulation~\cite{GPU-MC-small-N-1, GPU-MC-small-N-2, GPU-MC-small-N-3}.  We have not found any previous realization of MC simulation for large-scale long range interacting systems on GPU.  

In this work, we develop an efficient GPU approach to realize the canonical Monte Carlo of systems with long range interactions. The fundamental idea behind this method is to update every particle once during one invoking of GPU kernel, and to use one (or a few) thread(s) to control one particle in a synchronous mode with coalesced memory access, i.e., to calculate its energy change and to attempt to move it. The same idea has been used in MD simulations in~\cite{MD-GPU-Gem3}, where each thread controls the evolution of one particle. 
\footnote{However, there are also important differences between these two algorithms. In MD, the only dependency is that the particle coordinates updates and the force computations should be fairly separated in time. To satisfy this dependency, two basic kernels are needed: one is for position updating and the other one is for force computation. Moreover, the force computations are independent so that they can be performed without any inter-thread communication. In comparison, the Hamiltonian change in our MC method is more complicated, due to complex and strong dependency in updating coordinates and energy calculation. In our implementation for MC, the updates of coordinates and the corresponding energy computation are carried out within one kernel and the order of computations and updates together with inter-thread communications are deliberately designed. 
}
In our case of MC simulation, each thread also has to take care of Monte Carlo trials and decisions.  In terms of the interaction table, this parallel metropolis scheme looks like a ``brush'', hence we call it {\it the Brush Metropolis Algorithm}.

This approach enhances temporal locality and thereby improves cache performance. We benchmark this code on a Tesla K20 GPU and find a remarkable 440-fold speedup compared to sequential codes on Intel Xeon E5-2670.  It is important to stress that this speedup is achieved {\it without} sacrifice of accuracy, since there is no approximation (such as truncated multipole expansion) used in the Brush Metropolis Algorithm.  Using this program, we carry out large-scale Monte Carlo simulations of primitive model electrolytes containing as many as $10^6$ ions and measure all pair correlation functions up to extremely high precision.  The radius of the simulation box is hundreds of Debye length, so that all boundary artifacts are completely screened.  Using this huge amount of data, we are able to measure precisely static linear response properties of the system, including the renormalized valences of constituent ions, the renormalized Debye lengths and renormalized dielectric constant.  Comparison of these renormalized parameters with their bare values clearly demonstrates that the statistical physics of concentrated electrolytes is beyond the classical Poisson-Boltzmann theory.  

The remaining of this paper is organized as follows: In section 2, we compare the feasibility of parallelism in random moving and sequential moving. In section 3, we discuss GPU implementation of sequential moving. We show the benchmark results and present simulation results for linear response properties of electrolytes in sections 4 and 5, respectively. Section 6 summarizes our conclusions and future research directions.

\section{Concurrency}
\label{sec:concurrency} 
\begin{figure}[t!]
\centering
\includegraphics[height=0.30\textwidth]{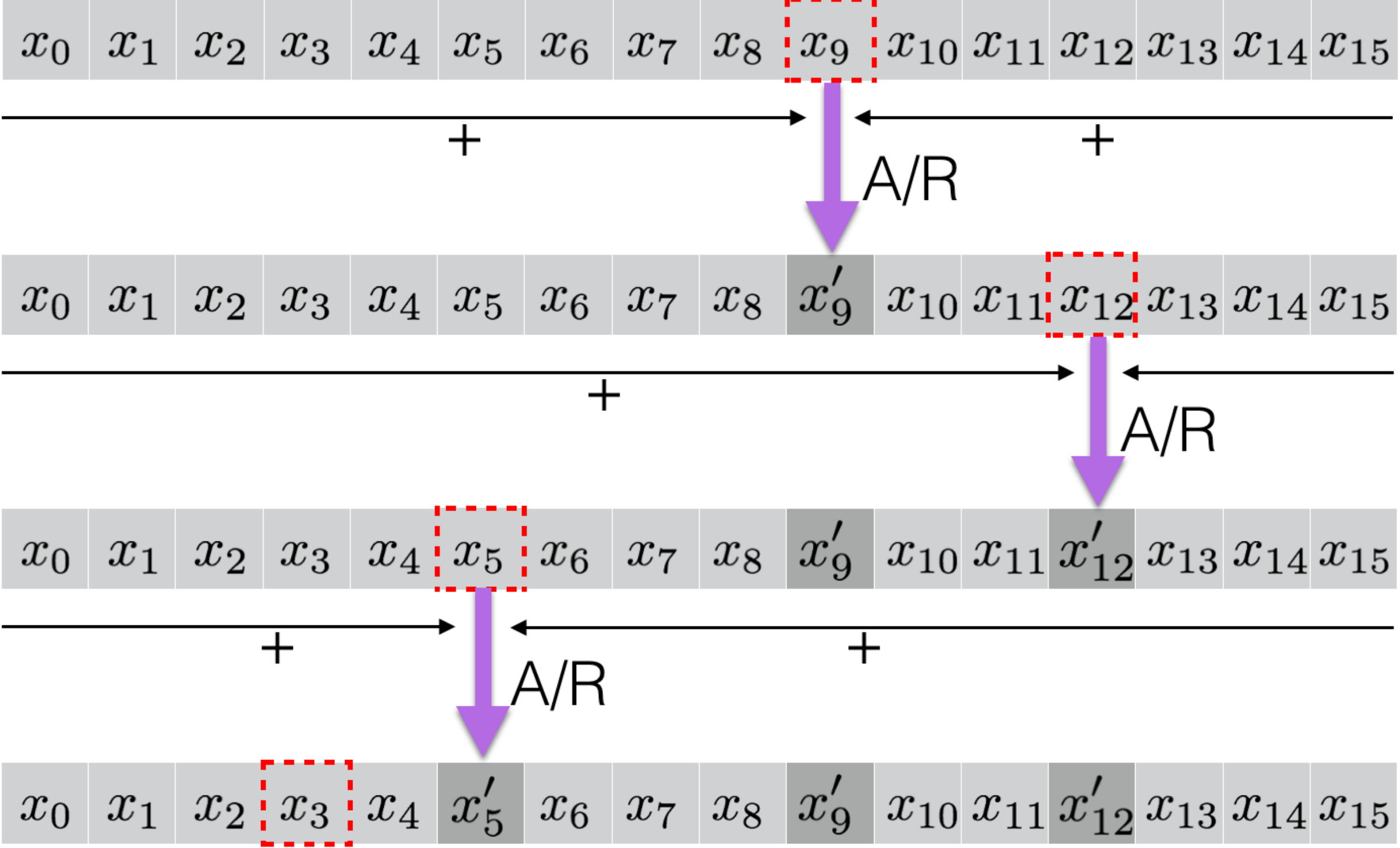} 
\includegraphics[height=0.30\textwidth]{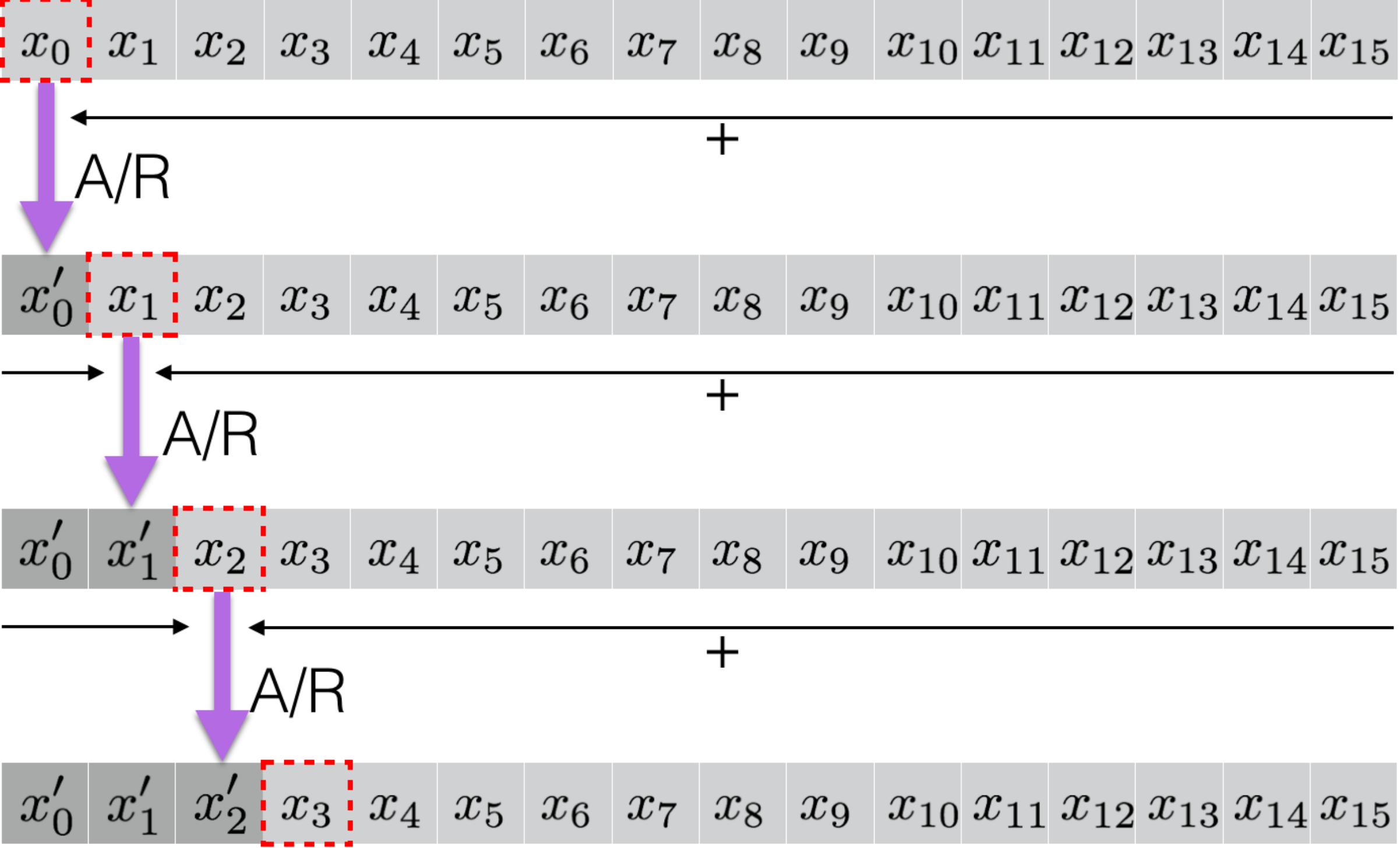}
\caption{Schematic illustrations of several consecutive MC steps in the random updating scheme (left) and the sequential updating orders (right).  Each row denotes a state of the system, and red dashed boxes denote the particles being updated, one in each step.  The purple arrows denote the actions of updating, with A/R meaning accept and reject, respectively.  In order to determine the results of updates (either acceptation or rejection), one needs to sum of the changes of all pairwise interactions. These are designated by black horizontal arrows with underlying plus signs.  }
\label{Update-Order}
\end{figure}

In a Metropolis Monte Carlo simulation, a Monte Carlo step is the smallest unit of Markov chain where one particle is moved.  A Monte Carlo cycle consists of $N$  steps, where $N$ is the total number of particles.   A MC step contains three basic sub-steps:
\begin{algorithmic}
\STATE a) {\bf Selection}: Select a particle $k$, either randomly or sequentially.

\STATE b) {\bf Trial}: Propose an unbiased perturbation of the selected particle. The new coordinate of the selected particle $\tilde{\xv}_k$ is generated by a symmetric probability transition function $T(\xv_k,\tilde{\xv}_k)=T(\tilde{\xv}_k,\xv_k)$, which yields conditional probability density that $\tilde{\xv}_k$ is selected as the new coordinate, given the current coordinate is $\xv_k$. We then calculate the change of total energy $\Delta E_k$ due to this trial. For long range interacting system, the time cost for calculation of $\Delta E_k$ scales with $N$, and therefore is the most computationally expensive substep.  

\STATE c) {\bf Acceptance/rejection}:  Accept the tried state as the next state of the Markov chain with probability ${\rm min}\{1,e^{-\beta\Delta E}\}$.  The new position is then given by
\be
\xv'_k = \left\{\begin{array}{ll}
\tilde{\xv}_k, & \mbox{trial accepted}; \\
\xv_k, & \mbox{trial rejected}. 
\end{array}
\right.
\label{trial-outcome}
\ee

\end{algorithmic}
Here $\beta=1/(k_BT)$, $k_B$ is the Boltzmann constant and $T$ is the temperature of the system.
Substep a) ({\rm selection}) can be executed in two different ways:  1) Random Updating Scheme, where the particle is selected at random, and 2)  Sequential Updating Scheme, where all particles are labeled and moved in ascending order.  In Fig.~\ref{Update-Order}, we schematically illustrate several consecutive MC steps in the random updating scheme (left) and the sequential updating orders (right).  All particles are labeled by integers $0,1,\ldots, N-1$.  Each row of figure corresponds to a state in one MC cycle, where the current position of each particle is listed in ascending order.  In the random scheme, every particle moves one step {\it on average} within one MC cycle. By contrast, in sequential updating scheme, every particle attempts to move exactly once within one cycle.  In random updating scheme, the Markov process has a time-independent transition matrix, whereas in sequential scheme, the transition matrix is periodic with periodicity $N$.   Note that detailed balance of the Markov process is guaranteed by the symmetry of the transition function $T(\xv, \tilde{\xv})$, together with the  choice of acceptance probability, and therefore is valid for both updating schemes. Because of the Markovian nature of MC simulation, the acceptance ratio of a trial in the random updating scheme depends on all details of all prior steps. This severely limits the potential of concurrency of the Random Updating Scheme.  By contrast, the potential of concurrency of Sequential Updating Scheme is much higher, as we shall show in detail. 

\begin{figure}[h]
\centering
\includegraphics[width=9cm]{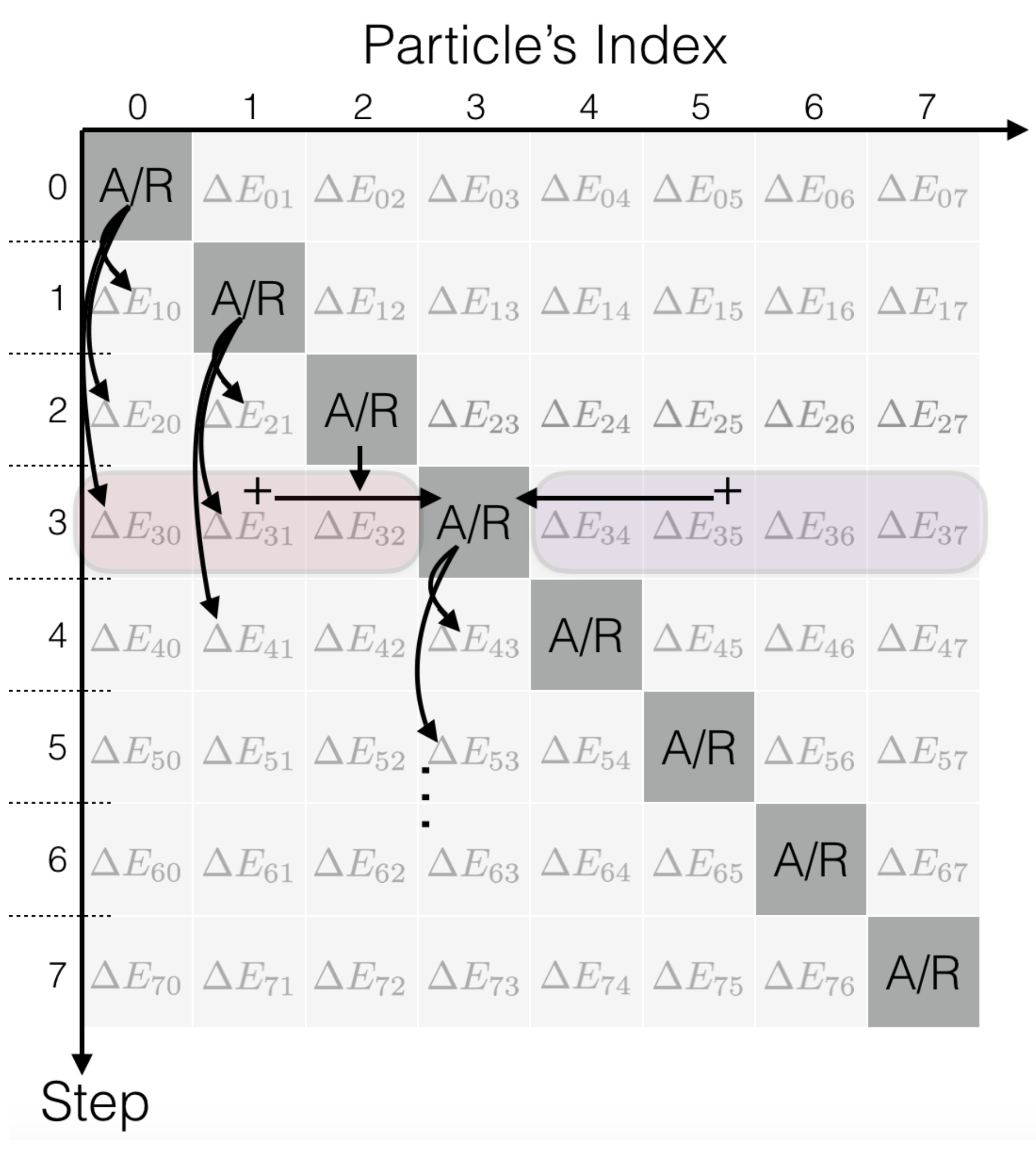}
\caption{Table of pairwise interactions in sequential updating scheme.  Each element (except the dark ones) denotes computation of a pairwise interaction. The dark square (one for each row) denotes the action of decision(acceptance or rejection) for the selected particle in each step. ``$+$'' denotes the summation of the pairwise interactions in selected region (red region or purple region). Decision must be done after the summations in each row. After the decision in one column is done, all squares below are to be updated. } 
\label{Table}
\end{figure}

Starting from the initial state $\{\xv_0, \xv_1, \ldots, \xv_{N-1}\}$, we can now propose $N$ random trial new positions $\{\tilde{\xv}_0, \tilde{\xv}_1, \ldots, \tilde{\xv}_{N-1}\}$, one for each particle, so that the first several states in one MC cycle are as shown in Fig.~\ref{Update-Order}. To illustrate the inter-dependency of tasks involved in one MC cycle of the sequential updating scheme, we introduce the graphic representations in Fig.~\ref{Table} of all involved pairwise interactions that must be calculated during a MC cycle.  The horizontal axis is the index of particles (in ascending order), whereas the vertical axis is the time line, with the numbers on the axis label particles selected in each step.  Each square element (except the dark ones) denotes the change of a pairwise interaction $\Delta E_{ij}$ in step $i$ where the $i$-th particle is being tried to move.  It is  defined as 
\be
\Delta E_{ij} \equiv \left\{ \begin{array}{ll}
U(\tilde{\xv}_i - \xv_j) - U(\xv_i - \xv_j), & i < j; 
\vspace{4mm}\\
U(\tilde{\xv}_i - \xv'_j) - U(\xv_i - \xv'_j), & i > j.
\end{array}
\right.
\ee 
Now the important point is that for $i<j$, $\Delta E_{ij}$ does not depend on any of new positions $\xv'_k$.  (All these $\Delta E_{ij}$ are shown in the upper triangle in Fig.~\ref{Table}. ) Therefore they can be calculated simultaneously before any particle is moved. Furthermore, all the squares in one row to the right of diagonal can be summed before any move.  Both of these calculations can be done in parallel.  

Now the $0$-th particle is ready to take the trial move, and determines its new position according to Eq.~(\ref{trial-outcome}). After this, all elements in the 0-th column below the diagonal line can be calculated independently (because these elements depend only on the new position $\xv'_0$, but not on others).  The inter-dependency of these operations is illustrated in Fig.~\ref{Table}.  They can also be added to $\Delta E_i$ independently.   After this, particle $1$ is ready to take the trial move.  The process keeps going until the last particle $N$ is tried.  This is the main idea behind our GPU implementation of the sequential updating scheme.  

\section{GPU implementation}
We implement the sequential updating scheme with the NVIDIA CUDA programming model and perform simulations and benchmarks on Tesla K20 GPUs. CUDA is a programming platform with extensions of C/C++, which provides a convenient way for parallel programming on NVIDIA GPU. An Nvidia Tesla K20 card includes 13 stream multiprocessor(SMX), each of which has 192 CUDA cores, result in totally 2496 CUDA cores and 1.17 Tflops double-precision peak performance or 3.52 Tflops single-precision peak performance. The size of global memory in a Tesla K20 GPU is 5GB, and in each SMX there is a 48KB shared memory. Details and terminologies of CUDA and GPU architecthure can be found in the CUDA-C programming guide \cite{CUDA-doc}.

\subsection{Data setting and random numbers}
We use CUDA's $float4$ data type for particle parameters, where $x,y,z$ components are positions and $w$ stores valence of each particle.
The information of particles makes up an $N$-elements array, which is stored in the device's global memory. In the following text, we denote this array by $X_{dev}$. Using $float4$ data type with aligned access allows coalesced memory access to the arrays of data in device memory, resulting in efficient memory requests and transfers \cite{CUDA-doc}. All the tasks in MC moves are performed in GPU, except for generation of random numbers.

In our program, random numbers are generated by the code ``ran4'' described in the book {\it Numerical recipes}~\cite{Numerical-Recipes}. This code generates single-precision floating-point numbers which are uniformly distributed in the interval $[0,1)$. We use two $N$-elements $float4$ arrays to store the random numbers used in an MC cycle, one of which is on the host and the other is on the device. The array of random numbers on device is denoted by ``rnum'' in the following text and pseudo-code. The $x,y,z$ components of an element in ``rnum'' are used to generate new coordinates whereas $w$ is for decision(acceptance or rejection) in each trial. The computational cost of random number generations can be effectively masked by MC cycles carried out on GPU. This is due to the fact that during the GPU performing a MC cycle, the random numbers for the next MC cycle can be generated simultaneously on CPU supported by concurrent execution between CPU and GPU in CUDA. We use CUDA's built-in function ``cudaMemcpy'' to upload the random numbers to the device. Function ``cudaMemcpy'' contains an intrinsic synchronization so that the system copies data only after the previous GPU job in the same stream finishes. Therefore there is no worry that the random number array changes while the GPU kernel is still using it.

Data sampling and analysis can be either on the device or on the host. If we want to put some analysis tasks in CPU cores, we can download this array by built-in functions such as ``cudaMemcpy'' from device as well.
\subsection{Decomposition}
When the host program invokes a kernel function of MC cycle, GPU constructs a one-dimensional grid with $B$ thread blocks. Each thread block is one-dimensional and contains $S=\lceil N/B \rceil$ threads. In the following text, we use a  two-element tuple $(bid,tid)$ to identify each thread, where $bid$ shows which block the thread locates in and $tid$ is the relative identity of this thread within the block $bid$.

Similarly, the array of particles is divided into $B$ groups, each with $S$ particles.
The trial of $p$'th particle in group $g$ is assigned to the thread $tid=p$ in block $bid=g$. In this article, we call the assigned particle the host-particle of corresponding thread. 
The mission of each thread is to compute the change of energy due to the trial of its host-particle and then decide the acceptance of the new move. At the beginning of kernel grid, each thread loads the information of its host-particle and random number vector from the global memory to its own registers. Then it computes the new position of its host-particle. The total energy change $\Delta E_{tid}$ of the host-particle is stored as a double floating-point variable in thread's registers, which can be initialized with 0 or the change of its self-energy (if any). The pseudo-code of initialization and the statements of symbols are shown in Algorithm 1.

\begin{algorithm}
\caption{Statement and initialization}
\begin{algorithmic}
\REQUIRE bid: the index of block
\REQUIRE tid: the index of thread in its block
\REQUIRE B: number of groups and the number of blocks
\REQUIRE S: number of particles(threads) per group(block)
\REQUIRE $X_{dev}$: Particles' list on the global memory
\REQUIRE rnum: Random number sequence on the global memory
\REQUIRE $D$: Range of one moving
\REQUIRE $\Delta E$: Total change of energy due to the trial of host-particle. It is double precision and located on each thread's registers.
\REQUIRE BlockState: An array with B elements on global memory which shows the state of each block. The initial value of each element is $0$. The element is $1$ if the corresponding block terminated and the coordinates of it's host particle group updated.
\REQUIRE $X$: a float4 vector whose x,y,z components are old coordinates of the host-particle and w component is the valence
\REQUIRE $X'$: a float4 vector whose x,y,z components are new coordinates of the host-particle and w is the random number for decision
\REQUIRE $Y$: a $float4$ array with B elements on the shared memory, which stores the coordinates of guest particles
\REQUIRE $\lambda$: Bjerrum length

\STATE
\STATE $X\leftarrow X_{dev}$[bid*S+tid]
\STATE $X'\leftarrow $rnum[bid*S+tid]
\STATE $X'$.x$\leftarrow (X'$.x $-0.5)*D$
\STATE $X'$.y$\leftarrow (X'$.y $-0.5)*D$
\STATE $X'$.z$\leftarrow (X'$.z $-0.5)*D$
\STATE $\Delta E\leftarrow$ Change of self energy
\end{algorithmic}
\end{algorithm}

\subsection{Energy computation}
Despite the positions are represented by single floating-point data type, all the computations related to the energy should be performed by double floating-point operations. The reason is that the energy is a summation of billions of terms which can be either positive or negative in the charged system. Single float treatment in such summation amplifies the truncation error significantly.

If there are hard core repulsions between particles, energy may be infinite. To avoid this problem, we can set an individual variable to indicate overlap. All the updating scheme of this indicator is similar to the computation of energy. In the CPU-based sequential program, when the overlap is encountered, we can stop the energy computation and reject the trial directly. This pre-rejection strategy can improve the speed of simulation on CPU, especially in the high volume fraction system. However, this is not a GPU-friendly strategy, due to the fact that this strategy leads to a long divergence path among different MC trials, which offends the SIMD computing scheme of GPU \cite{MC-EL} and degrades the parallel efficiency.
In this work, we do not use the pre-rejection strategy for two reasons. Firstly, the purpose of this work is to give a general framework on how to parallelize the Metropolis Monte Carlo simulation, where the hard-core repulsion is just a specific choice to deal with the short-range interaction. Secondly, in almost all the systems we study in this work, the volume fraction of particles is low, so that the pre-rejection doesn't make significant reduction of the computation, and the probability that all threads within a warp be pre-rejected is very low. As a result, the naive pre-rejection strategy on the GPU implementation cannot improve the performance.

\subsection{Intergroup Calculation}
When threads in a block compute interactions with group $J$ where $J\neq bid$ (particles in $J$ are called the {\bf guest particles} and $J$ is called the {\bf guest group}), they perform a brush-like operation which is widely used in MD implementation~\cite{MD-GPU-Gem3}. Threads in this block firstly load the information of $J$ from the global memory to an $S$-element float4 array $Y$ on the shared memory. This loading procedure is one-to-one and aligned, so that we can enhance the cache hit rate and reduce the transaction of global memory.

\begin{figure}[tbph]
\centering
\includegraphics[width=0.45\textwidth]{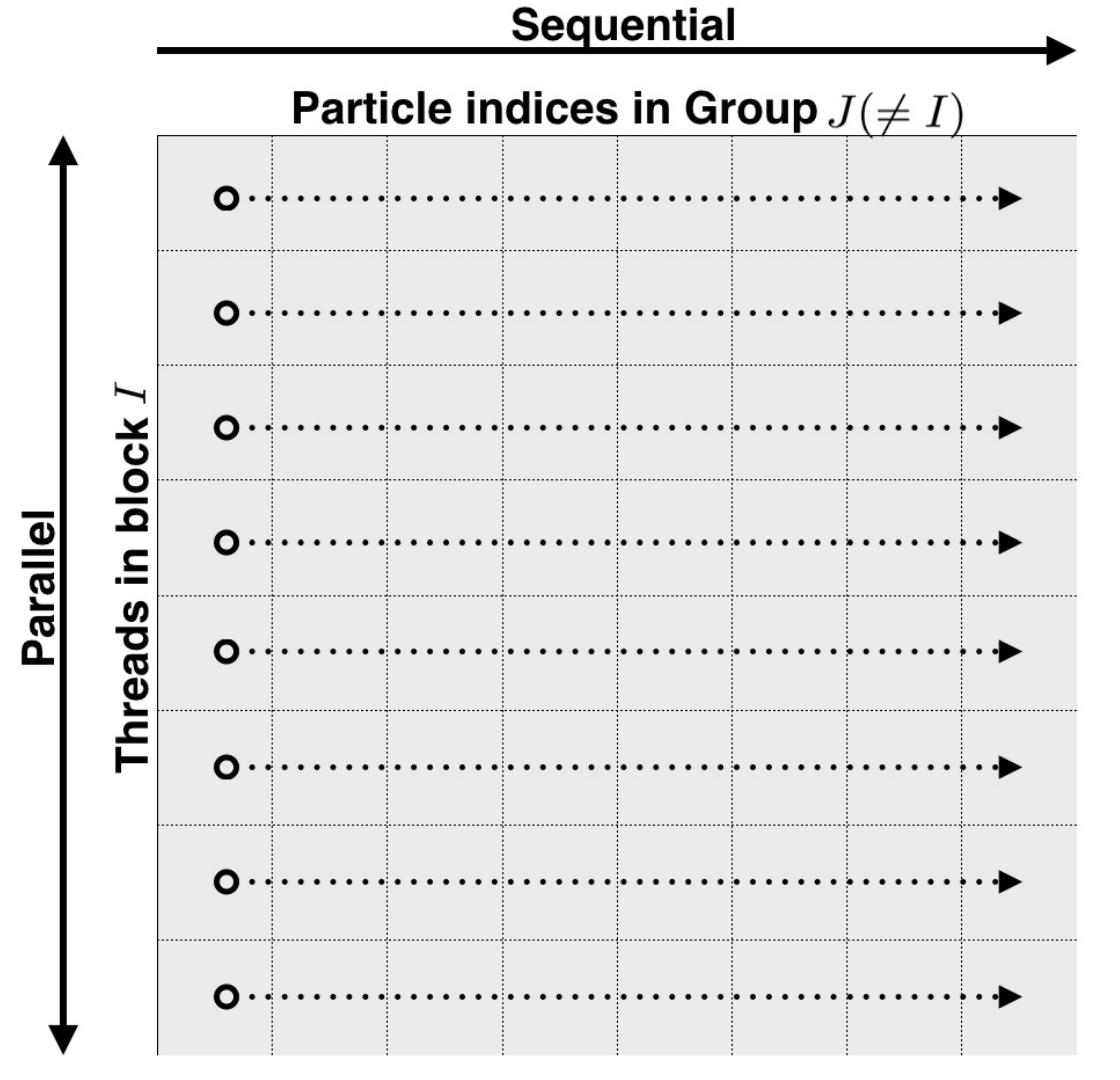}
\caption{Schematic illustration of intergroup calculation. In this interaction table, the vertical indices indicate threads in thread block $I$ and horizontal indices specify the particles in group $J$($\neq I$). Each element stands for a pairwise energy change of corresponding particle pairs. The dashed arrows show the direction of loop for summation. Threads in block $I$ compute and accumulate these pairwise interactions along horizontal direction in sequential order, whereas the energy change on each particle in group $I$ is evaluated in parallel. }
\label{Inter_Group}
\end{figure}

After group $J$ loaded (here a synchronize instruction ``\_\_syncthreads''  is taken to ensure that all the threads in this block finish loading data of guest-group $J$), all threads within this thread block take a loop to calculate the change of pairwise interactions and add them up, as showed in Fig.~\ref{Inter_Group}.

To avoid the bank conflict and take advantage of broadcasting feature of shared memory, all the loops in this block start at the same index. In this way, threads in the same warp access the same address simultaneously so that the shared memory can broadcast data. We choose 64-bit mode to make full use of the band-width of the shared memory.

At last, the block synchronize instruction ``\_\_syncthreads'' is taken again to ensure that all threads finish processing the group $J$ before the next group are loaded. Fig.~\ref{Inter_Group} is the schematic illustration of this part. The pseudo-code of this part is shown in algorithm 2.

\begin{algorithm}
\caption{Energy changes with group J (J$\neq$bid)}
\begin{algorithmic}
\STATE Y[tid]$\leftarrow$ $X_{dev}$[J*S+tid];
\STATE \_\_syncthreads();
\FOR {$l:=0$ \textbf{to} $S-1$}
          \STATE $\Delta E\leftarrow \Delta E+\Delta U$($X$,$X'$,Y[$l$]);
\ENDFOR
\STATE \_\_syncthreads();
\end{algorithmic}
\end{algorithm}

\subsection{Self Calculation}
\begin{figure}[tbph]
\centering
\includegraphics[width=0.85\textwidth]{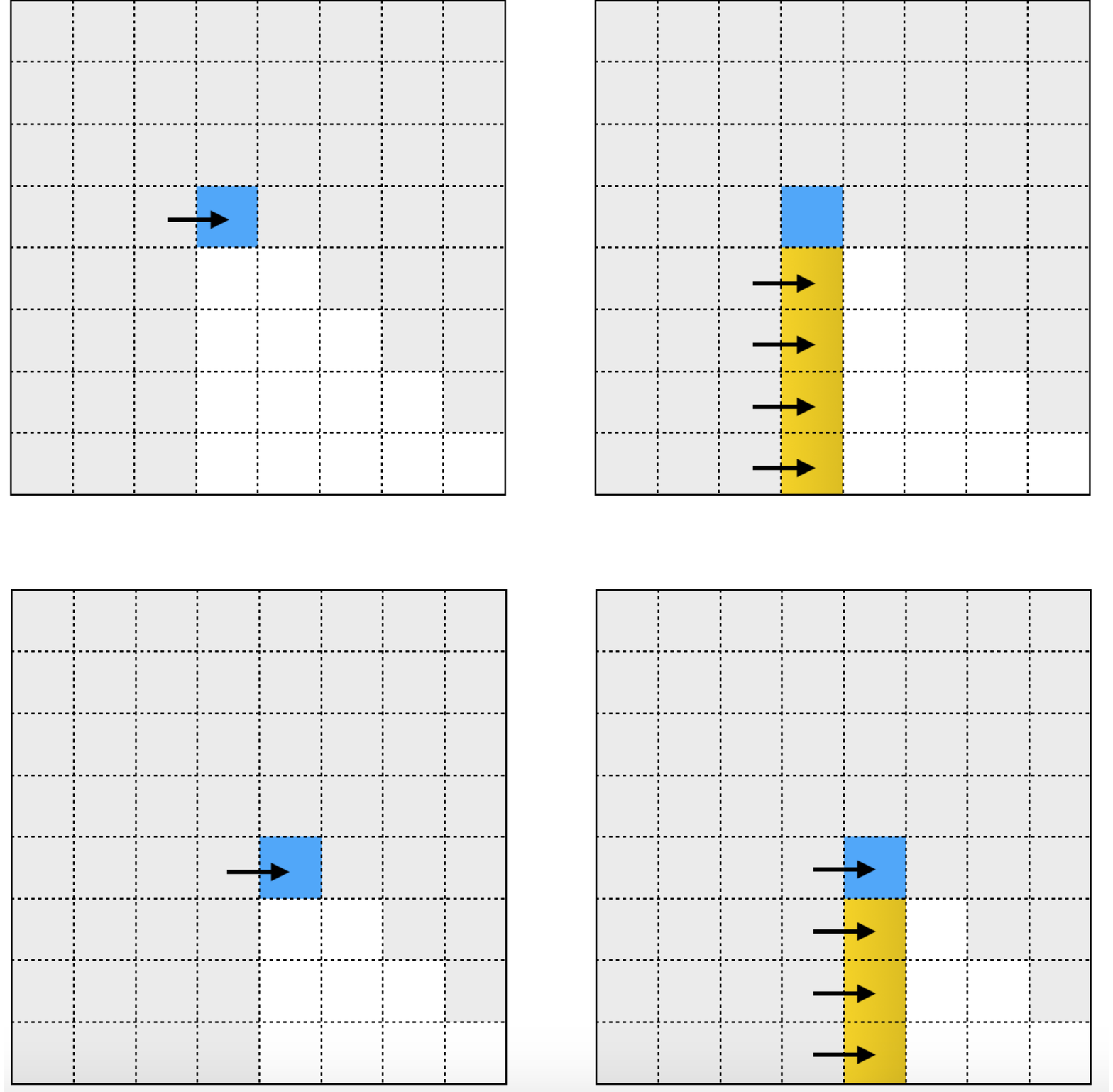}
\caption{Lower triangle part of self calculation. }
\label{Self_Block2}
\end{figure}

When thread block computes the interactions within its host-group, the procedure involves two main tasks of computations: (1). Compute the upper triangle parts in Fig.\ref{Self_Block1}. (2). Decide the trial and calculate the lower triangle parts as in Fig.\ref{Self_Block2}. Here we describe the details of these two tasks respectively.

\subsubsection{Upper triangle part}
\begin{figure}[tbph]
\centering
\includegraphics[width=0.85\textwidth]{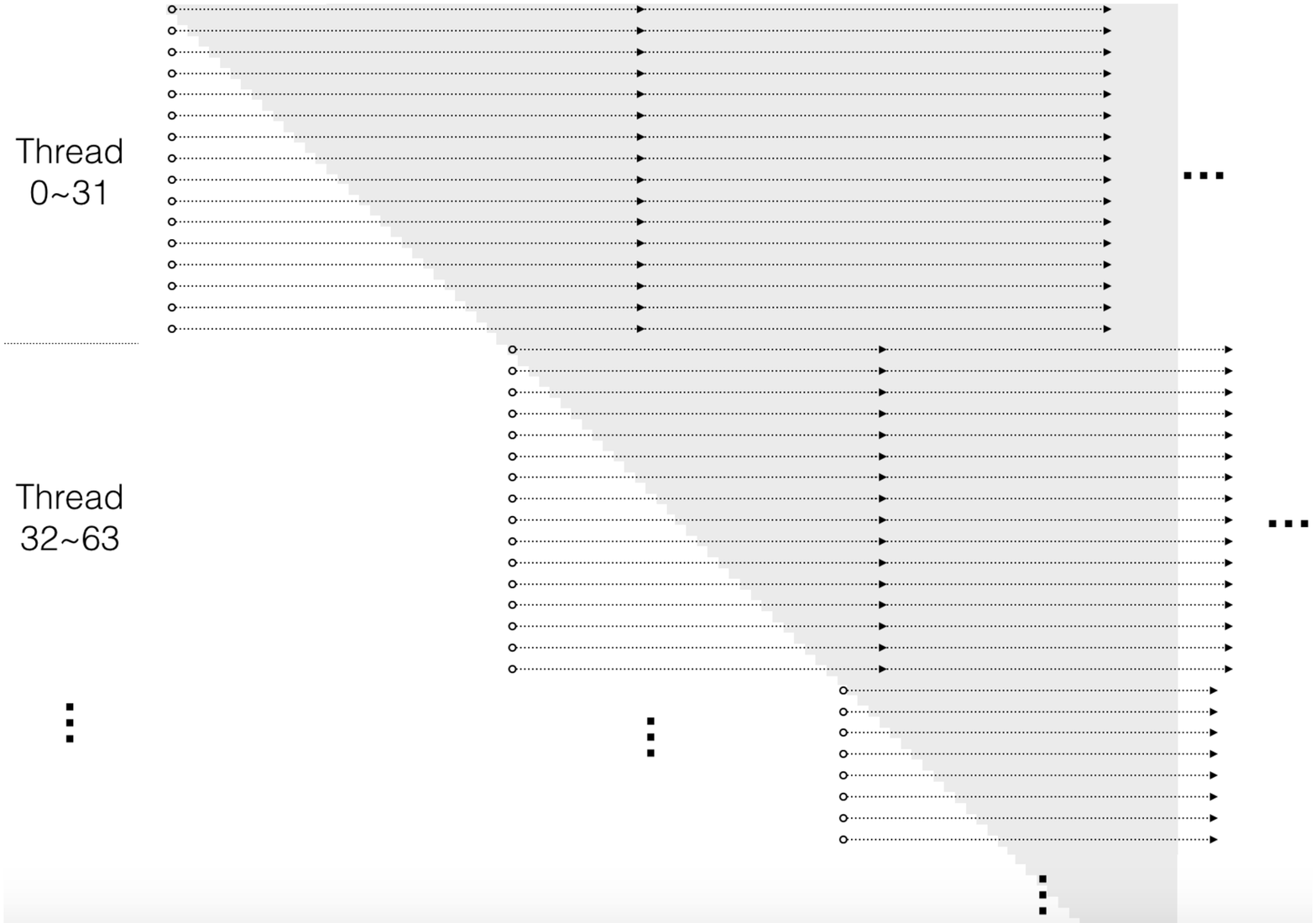}
\caption{Upper triangle part of self calculation. }
\label{Self_Block1}
\end{figure}

The upper triangle of the self-interaction table is independent of any other tasks. Therefore we can pre-calculate them by the brush scheme as before. Thread $tid=p$ computes the change of energy with particles $p+1$, $p+2$, ..., $S-1$ in the host-group. To ensure threads within a warp accessing the same address, the for-loop of each thread should start at $\lfloor tid/w\rfloor \times w+1$. Here $w$ is the size of a warp. In most GPU, $w=32$. Fig.~\ref{Self_Block1} is an illustration of this procedure.
The pseudo-code for this part is shown in algorithm 3. 
\begin{algorithm}
\caption{Upper triangle part of self calculation}
\begin{algorithmic}
\STATE Y[tid]$\leftarrow$ $X$;
\STATE \_\_syncthreads();
\FOR {$l:=$$\lfloor tid/w\rfloor \times w+1$ \textbf{to} $S-1$}
   \IF{$l>$tid}
          \STATE $\Delta E\leftarrow \Delta E+\Delta U$($X$,$X'$,Y[$l$]);
   \ENDIF
\ENDFOR
\STATE \_\_syncthreads();
\end{algorithmic}
\end{algorithm}

In the first $w$ loops, half of the threads on average within a warp are inactive. To reach the full warp efficiency we can use the {\bf map algorithm} \cite{Energy-evaluation-GPU,Three-body-GPU,GPU-MC-large-N-1}. But this strategy increases usage of registers per thread that limits the number of concurrently executed blocks in one streaming multiprocessor(SMX) and thus decreases the SMX's efficiency. On the other hand, the computation of the first $w$ loops is light compared with the whole tasks. So we do not implement this strategy in our code for Tesla K20.

\subsubsection{Decision and lower triangle part}
The computation of the lower triangle part of the self-interaction table, together with decision of the trial in the host-group, should be performed at last. As shown in Fig.~\ref{Self_Block2}, when thread $0$ finishes all the summations for $\Delta E_0$, it can make a decision by comparing with the random number $X'$.w under acceptance ratio $\min\{1,\exp(-\beta\Delta E_0)\}$. If $X'$.w$<\min\{1,\exp(-\beta\Delta E_0)\}$, thread $0$ replaces $\xv_0$ by $\xv'_0$ on the shared memory. Otherwise no coordinate update will be made. After thread $0$ updates the coordinate of $0$'th particle, the remaining threads $tid>0$ in this block calculate the interactions with the $0$'th particle together, and add the results to their own $\Delta E_{tid}$. Then thread $1$ updates the coordinate of particle $1$ and the updated energy of particle $1$ is propogated to the other threads. This procedure is repeated for other threads until all the host-particles in this block are updated. 

The last step is to copy the information array of the host-particles from the shared memory to the global memory and set the corresponding flag ``BlockState'' to be one. The ``BlockState'' is an array with $B$ elements, each of which indicates the state of corresponding block. Initially all the elements in this array is $0$. When a thread block has written the updated coordinates of its host-particles to the global memory, the corresponding flag BlockState[$bid$] is set to $1$. Since CUDA is a weak order programming language, the writing instruction for coordinates and the setting instruction for ``BlockState'' should be seperated by a ``syncthreads''or a ``threadfence'' instruction to ensure the required executing order viewed in other blocks. The pseudo-code for this part is shown in algorithm 4.

\begin{algorithm}
\caption{Decision and lower triangle part}
\begin{algorithmic}
\STATE Y[tid]$\leftarrow$ $X$;
\STATE \_\_syncthreads();
\FOR {$l:=0$ \textbf{to} $S-1$}
      \IF{(tid$=l$) and ($X'$.w$<\exp(-\lambda\Delta E)$)}
           \STATE $X'$.w$\leftarrow$$X$.w;
           \STATE Y[$l$]$\leftarrow$$X'$;
      \ENDIF
      \STATE \_\_syncthreads();
      \IF{(tid$>l$)}
          \STATE $\Delta E\leftarrow \Delta E+\Delta U$($X$,$X'$,Y[$l$]);
      \ENDIF
\ENDFOR

\STATE $X_{dev}$[bid*S+tid]$\leftarrow$ Y[tid];
\STATE \_\_syncthreads();
\STATE Thread 0 do: Set BlockState[bid] to be 1
\end{algorithmic}
\end{algorithm}

\subsection{Global procedure}
\begin{figure}[tbph]
\centering
\includegraphics[width=1\textwidth]{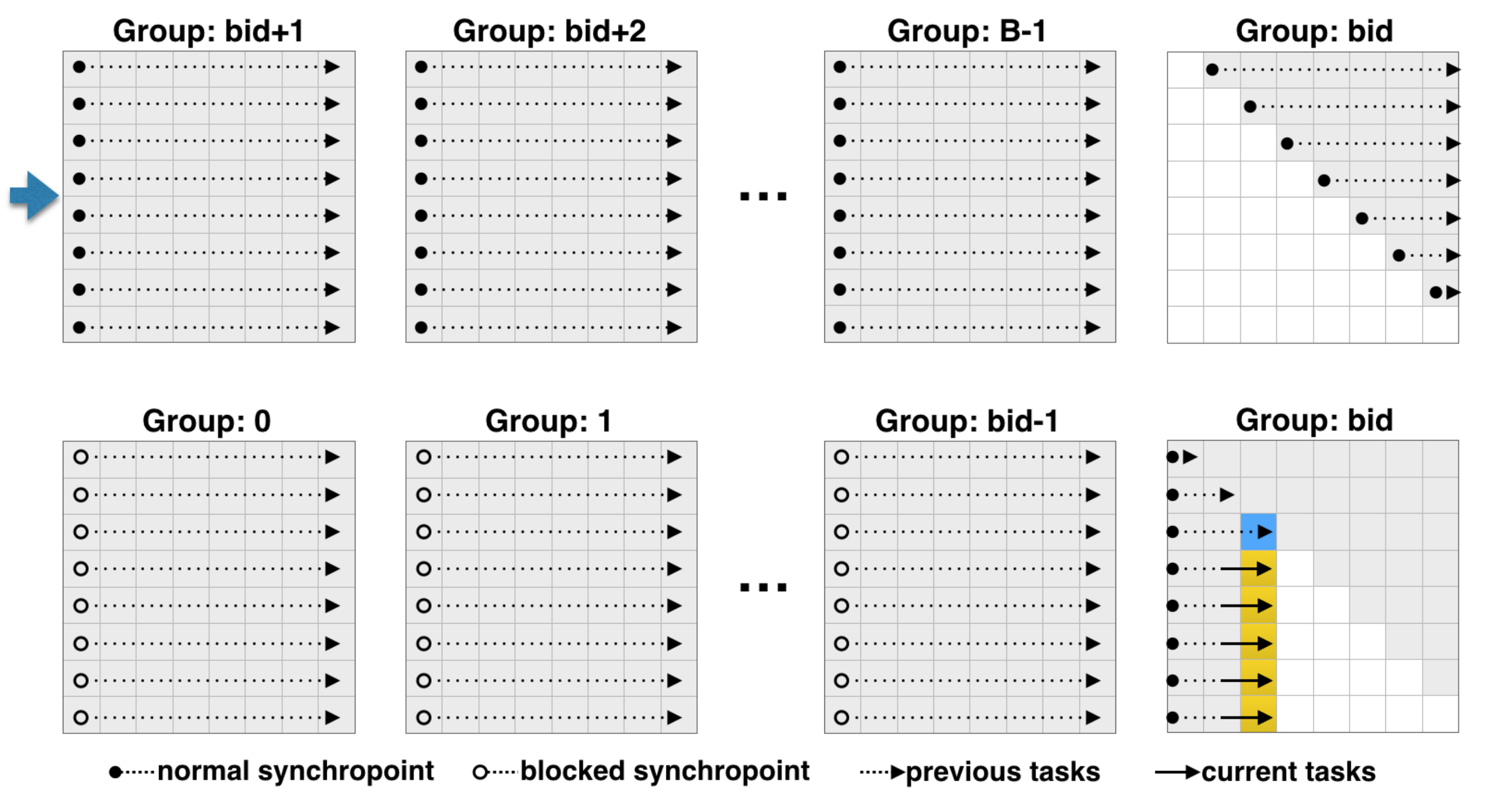}
\caption{Global procedure in Monte Carlo kernel. }
\label{Global_Scheme}
\end{figure}

Fig.~\ref{Global_Scheme} shows the global procedure of a thread block. After initialization, threads in this block firstly compute the inter-group interactions with groups $bid+1$,$bid+2$,\ldots$,B-1$. Then they compute the internal upper-triangle interactions.

The second global loop is to evaluate the interactions with groups $0$,$1$,..,$bid-1$. In serial implementation, particles in these groups are updated before the host-particles. To realize this dependency in the GPU implementation, before loading information of a guest-group $J$, $J<bid$,  all threads in this block should wait until the coordinates of particles in guest-group $J$ are updated. Therefore, in each step $J$ of this loop, one of the threads in this block firstly runs a subloop to check and wait until BlockState[$J$] is 1. Finally, threads make the decisions and evaluate the lower-triangle part of self-interaction table.

A remaining issue is to reinitialize the flag-array to $0$s for the next grid. We assign this additional work to block $B-1$, which is the last one to terminate. Threads in it set all the elements of ``BlockState'' to $0$s so that the next grid can use these indicators directly.   The pseudocode for the global procedure is shown in algorithm 5.

\begin{algorithm}
\caption{Global procedure}
\begin{algorithmic}
\STATE{Initialization}
\FOR{$J:=$bid+1 \textbf{to} $B-1$}
 \STATE {Calculate the energy changes with group $J$}
\ENDFOR
\STATE {Calculate the energy changes of upper triangular part}

\FOR{$J:=0$ \textbf{to} bid-1}
	\STATE {Thread 0 do: Wait and check until BlockState[$J$]$=$1}
	\STATE {\_\_syncthreads();}
	\STATE {Calculate the energy changes with group $J$}
\ENDFOR

\STATE {Decision and lower triangular part}

\IF {bid$=$B-1}
\STATE {Set all the elements of BlockState to 0}
\ENDIF
\end{algorithmic}
\end{algorithm}

The above inter-block communications via flags are {\bf blocked communications} since receivers will not carry out follow-up tasks until receiving the required messages. In a parallel program with blocked communication, we always need to avoid a deadlock state where receivers are waiting for each other and cannot terminate without external force.
A case of deadlock in GPU is that all the resident blocks in SMXs are waiting for a message from a block which is not executed yet. In this case, the inactive sender block cannot move forward until one of the resident blocks terminates so that there is enough space to launch this sender block. Therefore, the resident blocks and the inactive sender blocks are waiting for each others.
In our implementation, deadlock can be fortunately avoided because each thread block receives messages only from its previous blocks.

\subsection{Multi-Thread-Per-Particle Treatment}
In above algorithm, one thread is responsible for one particle's trial. This one-thread-per-particle assignment cannot take full use of device if the number of particles is small because insufficient number of threads are carried out simultaneously to fully take advantage of the capability of GPU. To deal with small system, we take a multi-thread-per-particle assignment, in which the change of energy due to the trial of one particle is evaluated by two or more adjacent threads. 

\begin{figure}[tbph]
\centering
\includegraphics[width=1\textwidth]{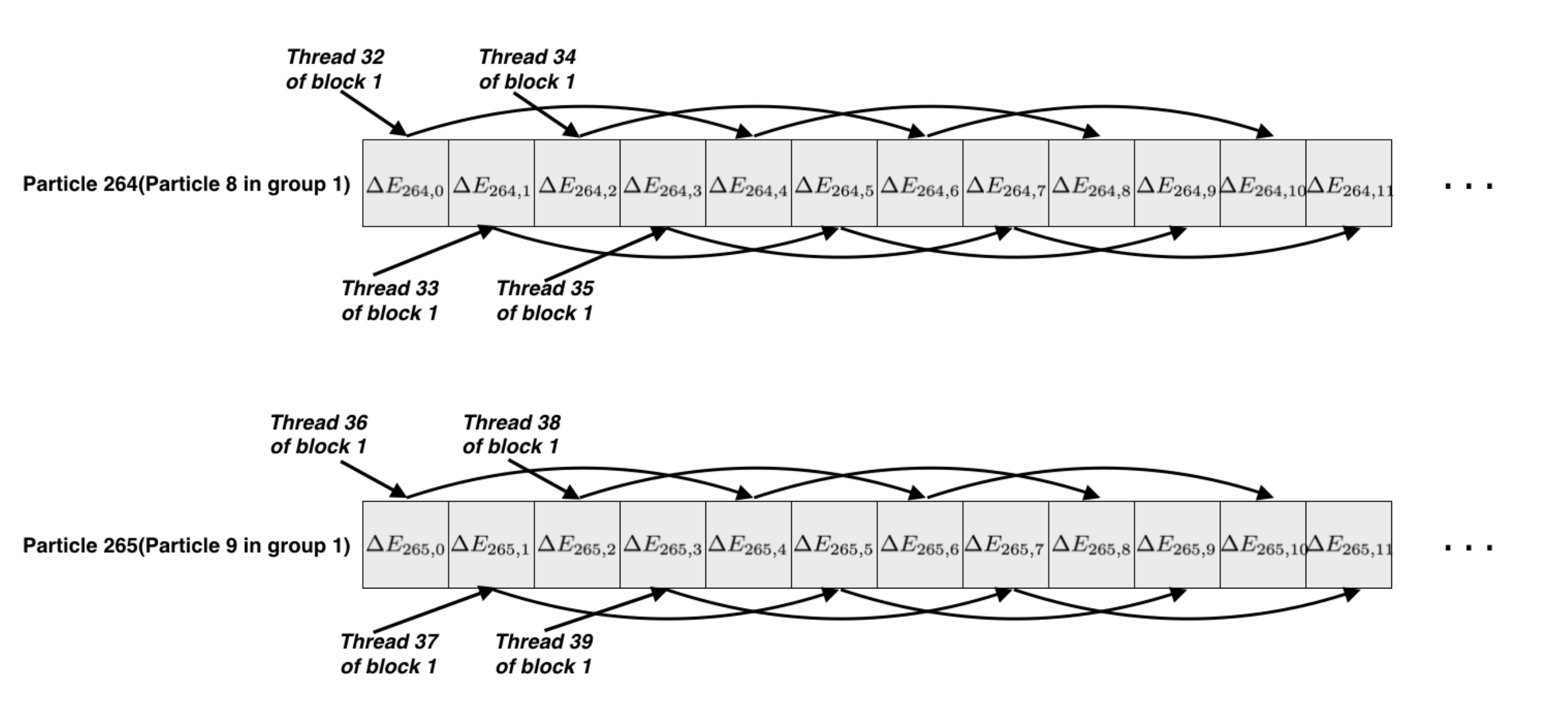}
\caption{Multi-thread-per-particle assignment for the intergroup evaluation. Each thread block contains 1024 threads. The size of particle group is 256.}
\label{MTPP}
\end{figure}

In a program with $\eta$-thread-per-particle scheme, the size of thread block is $\eta$ times of the group size. The energy change of the host particle $k$ ($k$ is the relative index within its group) will be computed by a bundle of threads $tid=\eta k$, $tid=\eta k$, ..., $tid=\eta (k+1)-1$, with interleaved loops as show in Fig.~\ref{MTPP}. This interleaved approach is performed in the inter-group evaluation and upper triangle part of self-table. Before the decisions and evaluation of the lower triangle part of self-table, we use warp-reductions to sum up results in all the slave threads of a particle to one. For convenience, $\eta$ is selected to be $2$, $4$, $8$ and $16$, leading $1$-step, $2$-step, $3$-step and $4$-step warp shuffle instructions in the reductions.

\section{Benchmark}
We benchmark the performance of our Brush Metropolis Algorithm on GPU by simulating the primitive model of electrolyte with various numbers of particles, and make a comparison with  two other MC implementations: a sequential CPU code and a GPU code with {\bf  parallel reduction scheme}.

In the CPU code, the evaluation and summation of the pairwise interactions are performed sequentially without using any acceleration methods. In the GPU code with parallel reduction scheme, each kernel tries to move one particle. There are $N-1$ pairwise interactions per trial to be calculated and summed, between the selected particle and the rest of particles in the system. We distribute these $N-1$ evaluation to $N$ threads while the thread for the selected particle remains idle. In this reduction kernel, the size of thread block is 1,024, which is the maximum number of threads per block in Tesla K20. The reduction within one block is done by warp shuffle instructions \cite{Reduction-Warp}, which is an optimized version of parallel reduction \cite{Reduction-Normal} for Tesla K20. The intermediate results of warp sums are stored in the shared memory temporarily.  We perform warp reduction again with these intermediate results  so that the partial sums in this block are integrated.  
The sub-total of each block is to be added into the global memory atomically. When a block finishes adding the sub-total to the global memory, it should increase a global counter by 1 atomically. This global counter indicates the number of completed thread blocks. Finally, the last block decides acceptance of the proposed particle move. Before the last block updates the coordinate of the selected particle, one of threads in the last block checks this global counter ceaselessly until it indicates  completion of summation.
This method is straightforward and easy to implement, but it has some drawbacks that limit its performance. Firstly, in the parallel reduction there is a great thread divergency. For a reduction within warp(32 threads), the mean number of active threads is only $(32+16+8+4+2+1)/6\approx 10.5$, which means about $2/3$ of computational resources are idle. 
Secondly, each time when the kernel is invoked, the system needs to take some time to initialize the environment.
In this naive method program invokes kernel for $N$ times in a cycle ($N$ is the number of particles) whereas our Brush Metropolis method just invokes kernel once per cycle. Thus this naive method accumulates a considerable time for initialization.
Finally, in each time when the kernel is invoked, the information of particles needs to be reloaded, leading to many global memory transactions and thus low cache efficiency.

For the Brush Metropolis GPU code, we mainly measure the performance of one-thread-per-particle approach, with number of particles from $8,192$ up to $1,048,576$. In addition, we carry out benchmarks of multi-thread-per-particle approach with number of particles from $8,192$ to $32,768$. We find that the size of block $1,024$ exhibits best performance. However, for the multi-thread-per-particle implementation, the number of slave threads per particle $\eta$ with best performance changes with the size of system. We measure the execution time for 4 small systems, and find the best $\eta$ for each system, as shown in Table.~\ref{Para}
\begin{table}[h]
    \begin{center}
    \begin{tabular}{ l | c | r }
    \hline
    $N$ & Size of thread block & $\eta$ \\ \hline
    8192 & 1024 & 16 \\ \hline
    16384 & 1024 & 8 \\ \hline
    24576 & 1024 & 4 \\ \hline
    32768 & 1024 & 4 \\ \hline
    \end{tabular}
    \end{center}
    \caption{Parameters}
    \label{Para}
\end{table}

We carry out all benchmarks on a Linux machine with $2\times 8$ Intel Xeon E5-2630 (2.3GHz) CPU cores and two NVIDIA Tesla K20 GPUs. The operating system is Ubuntu 14.04, with the host C-code compiler GCC 4.8.4 and the GPU code compiler CUDA 7.0. The GPU driver's version is 346.96.  The Monte Carlo simulation program on CPU is compiled with optimization option -O3.  Both GPU codes are compiled with option -arch=sm\_35.  Since the GPU implementation of the Brush Metropolis method costs lots of registers, we take option -maxrregcount=65 as the register usage per thread to yield the best performance.

\begin{figure}[tbph]
\centering
\includegraphics[width=0.9\textwidth]{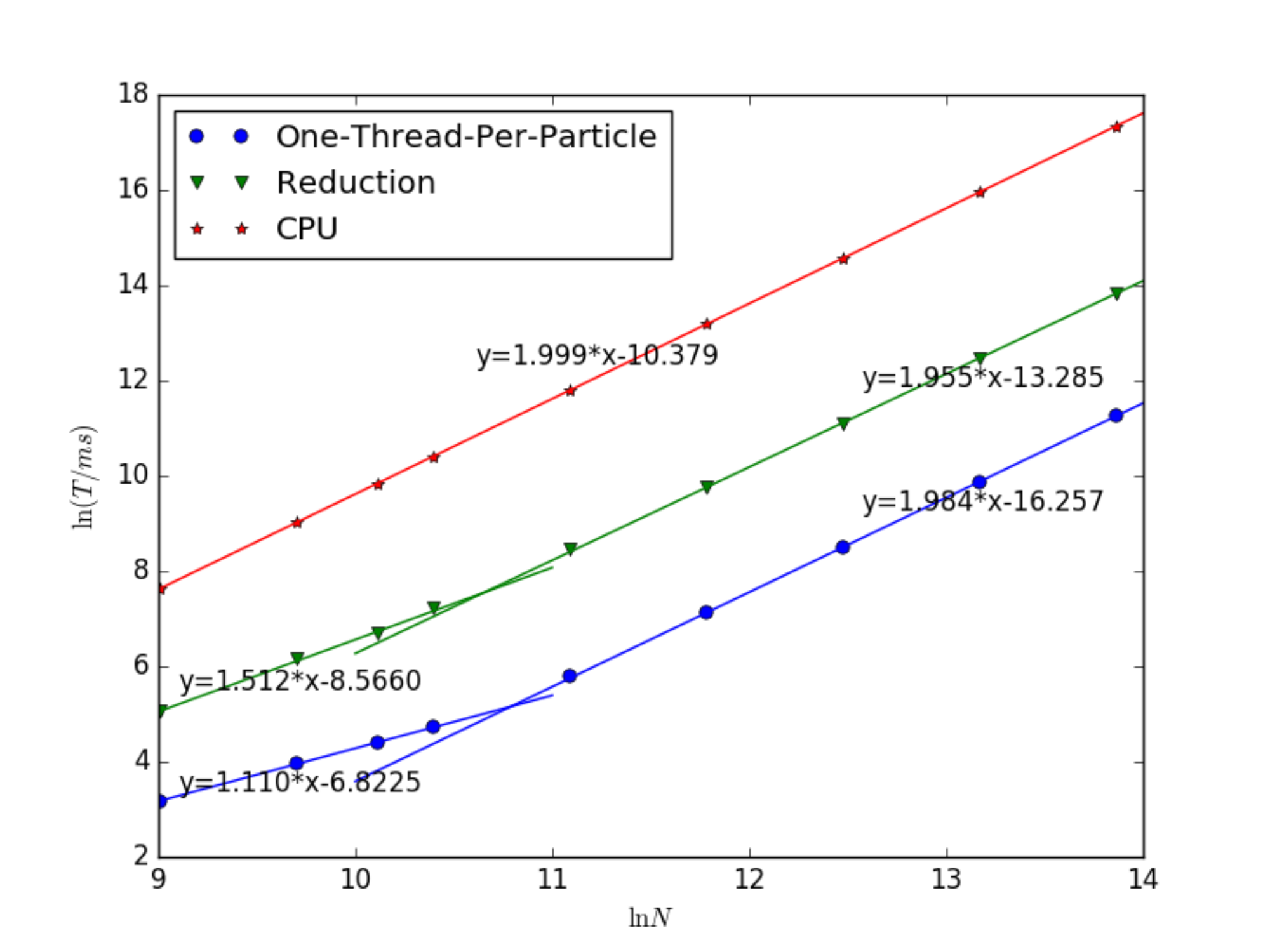}
\caption{Excution time in log-scale. }
\label{Exc-Time}
\end{figure}
\begin{figure}[tbph]
\centering
\includegraphics[width=0.7\textwidth]{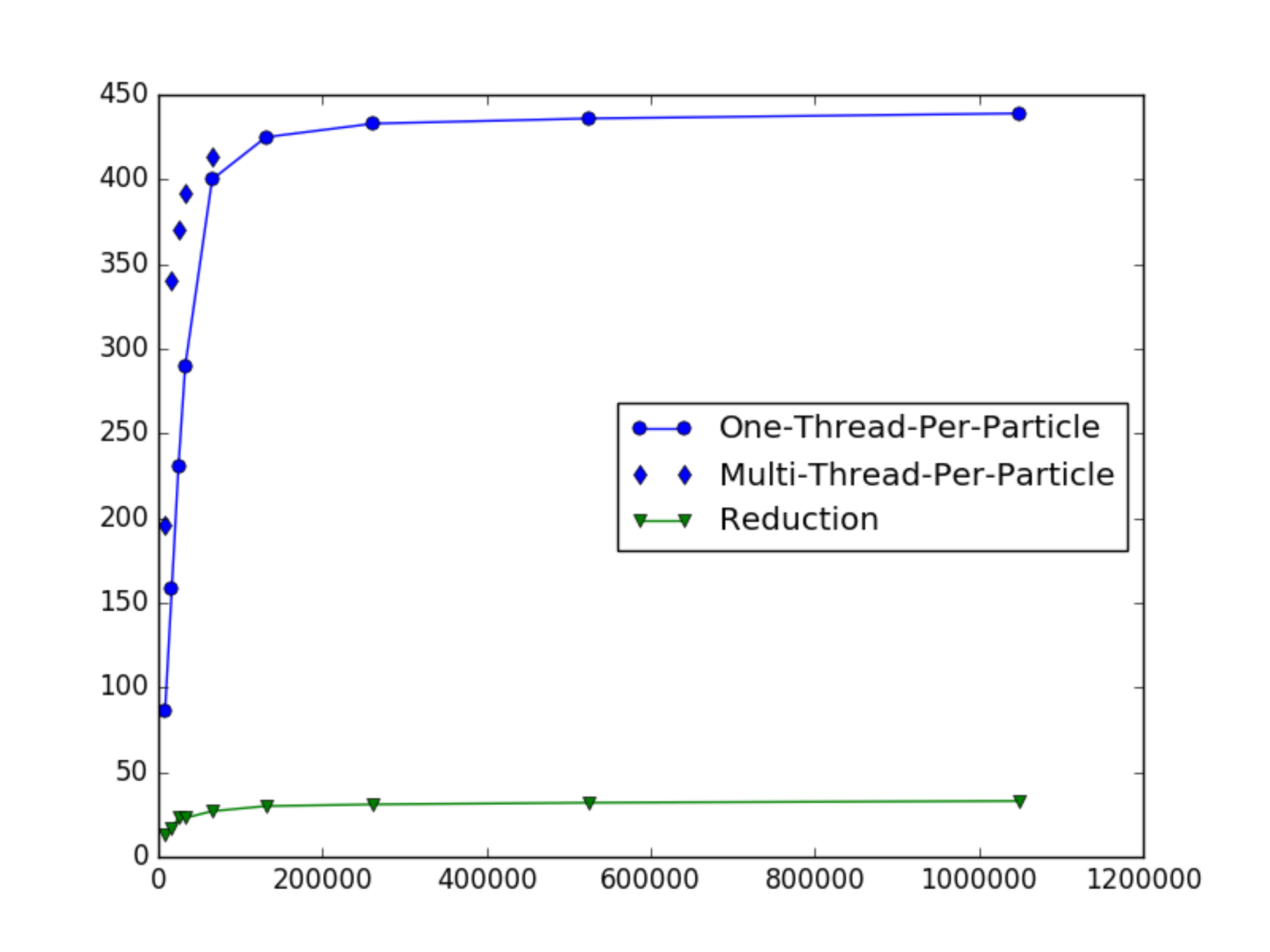}
\caption{Speedup of the Brush Metropolis GPU code and the convention GPU code using parallel reduction. }
\label{SpeedUp}
\end{figure}

\begin{figure}[tbph]
\centering
\includegraphics[width=0.8\textwidth]{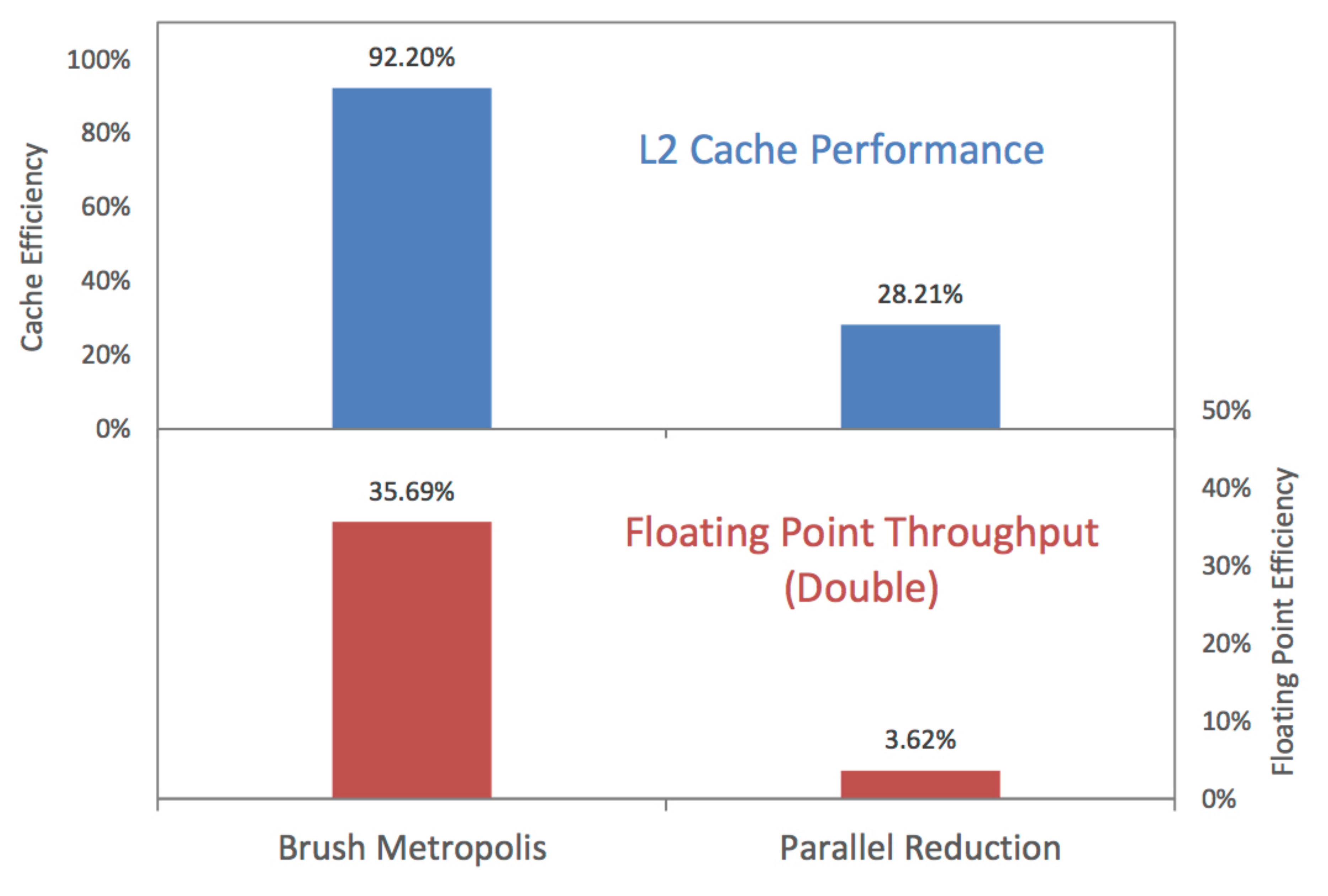}
\caption{Performance comparison between GPU implementations based on parallel reduction and Brush Metropolis on a system with 131,072 particles. Performance data obtained from nvprof. Due to significantly higher L2 cache efficiency, our GPU implementation has better floating point throughput than that of parallel reduction implementation.}
\label{Cache_Flop_Eff}
\end{figure}

The execution time in log-scale for all three algorithms is showed in Fig.\ref{Exc-Time}. It clearly shows that the time complexity of the Brush Metropolis Algorithm with one-thread-per-particle assignment is between the step complexity $O(N)$ and the work complexity $O(N^2)$. (The step complexity measures the computational complexity assigned to each thread, including the cost of synchronization.)  For particle number less than $50,000$, the time complexity scales approximately linearly with the step complexity. For even larger particle numbers, the workload of device is nearly saturated so that the time complexity scales with the work complexity. The GPU implementation with parallel reduction has a time complexity between $O(NlogN)$ and $O(N^2)$. 

The speedup of the Brush Metropolis GPU code (comparing with the CPU code) as a function of system size is showed in Fig.\ref{SpeedUp}.  For large systems the implementation with one-thread-per-particle saturates to approximately $440$. 
In general the Brush Metropolis method exhibits significantly better speedup than the implementation with parallel reduction method, because of less global memory transactions and higher L2 cache efficiency, as well as higher floating point throughput in comparison with parallel reduction method, as shown in Fig.\ref{Cache_Flop_Eff}. 
The main factor that prevents us from achieving even higher floating point efficiency is the Metropolis-style acceptance and rejection, which generate a divergence path for each proposed move.

\section{Simulation of Linear Response Properties of Dense Electrolytes}
\label{sec:simulation-LRT}
In this section, we use our Brush Metropolis GPU code to carry out a large scale MC simulation of the {\it primitive model} electrolytes, which are modeled as charged hard spheres.  In the experiment the system contains $1,048,576$ ions, and more than a hundred of Debye length. Because of the huge amount of available data, we are able to measure all pair correlation functions up to {\it extremely precision and very long scale}.   Using these correlation functions, we determine various renormalized parameters that characterize the linear response properties of the electrolyte, including renormalized valences of the constituent ions, the renormalized Debye length, and the renormalized dielectric constant.  These results demonstrate unequivocally that the properties of system are beyond the classical Poisson-Boltzmann theory (PB).  We emphasize that in order to compute precisely the long scale properties of electrolytes, it is essentially important to simulate very large system sizes {\it without making any approximation in long scales}. This is the main advantage of our GPU code, compared with methods using multipole expansions. 

\subsection{Charge renormalization in concentrated electrolytes}
The classical Poisson-Boltzmann theory (PB) \cite{Andelman-PBequation} predicts that the mean potential around a fixed ion with charge $Q$ is given by
\begin{equation}
\phi=\frac{Q \, e^{-\kappa r}}{4\pi\epsilon r},
\end{equation}
whereas the effective interaction, i.e., two-ion potential of mean force (PMF), between two ions $Q_1,Q_2$ is 
\begin{equation}
U_{12} = \frac{Q_1 Q_2 \, e^{-\kappa r}}{4\pi\epsilon r}.
\end{equation}
In the above equations, $\kappa$ is the {\it bare} inverse Debye length, which, according to PB, is related to bulk ion densities $\bar{n}_{\pm}$ and charges $q_{\pm}$ via:
\be
  \kappa^2 = 
 \frac{\beta}{\epsilon} 
\left(  \bar{n}_{+}q_{+}^2
+ \bar{n}_{-}q_{-}^2
\right).  
\label{alpha-kappa_0}
\ee

Because the classical PB theory ignores correlation effects, it is not applicable in the concentrated regime.  In this regime, the correct far field behaviors of mean potential and two-ion PMF are \cite{Kjellander:1992nr,DLX-ion-specific}:
\ba
\phi &=& \frac{Q_R \, e^{-\kappa_R r}}{4\pi\epsilon_R r},
\label{phi-DIT}
\\
U_{12} &=& \frac{Q^R_1 Q^R_2 \, e^{-\kappa_R r}}{4\pi\epsilon_R r},
\label{U_12-DIT}
\ea
where $Q_R, Q_1^R, Q_2^R, \kappa_R, \epsilon_R$ are, respectively, the renormalized charges, renormalized inverse Debye length, and renormalized dielectric constant, which are different from their bare values. There is an exact relation between the renormalized Debye length and renormalized charges of constituent ions:
\begin{equation}
\left(\frac{\kappa_R}{\kappa} \right)^2 ={\frac{q_+^R-q_-^R}{q_+-q_-}}. 
\label{RD}
\end{equation}
Eqs.~(\ref{phi-DIT})-(\ref{RD}) are the main results of the {\it dressed-ion theory}~\cite{Kjellander:1992nr}. 
We shall compute renormalized valences of constituent ions, renormalized Debye length and renormalized dielectric constant, and finally verify the relation Eq.~(\ref{RD}) using large-scale MC simulations.  

\subsection{Simulation Methodology}
\label{app:simulation}

\begin{figure}[tbph]
\centering
\includegraphics[width=0.8\textwidth]{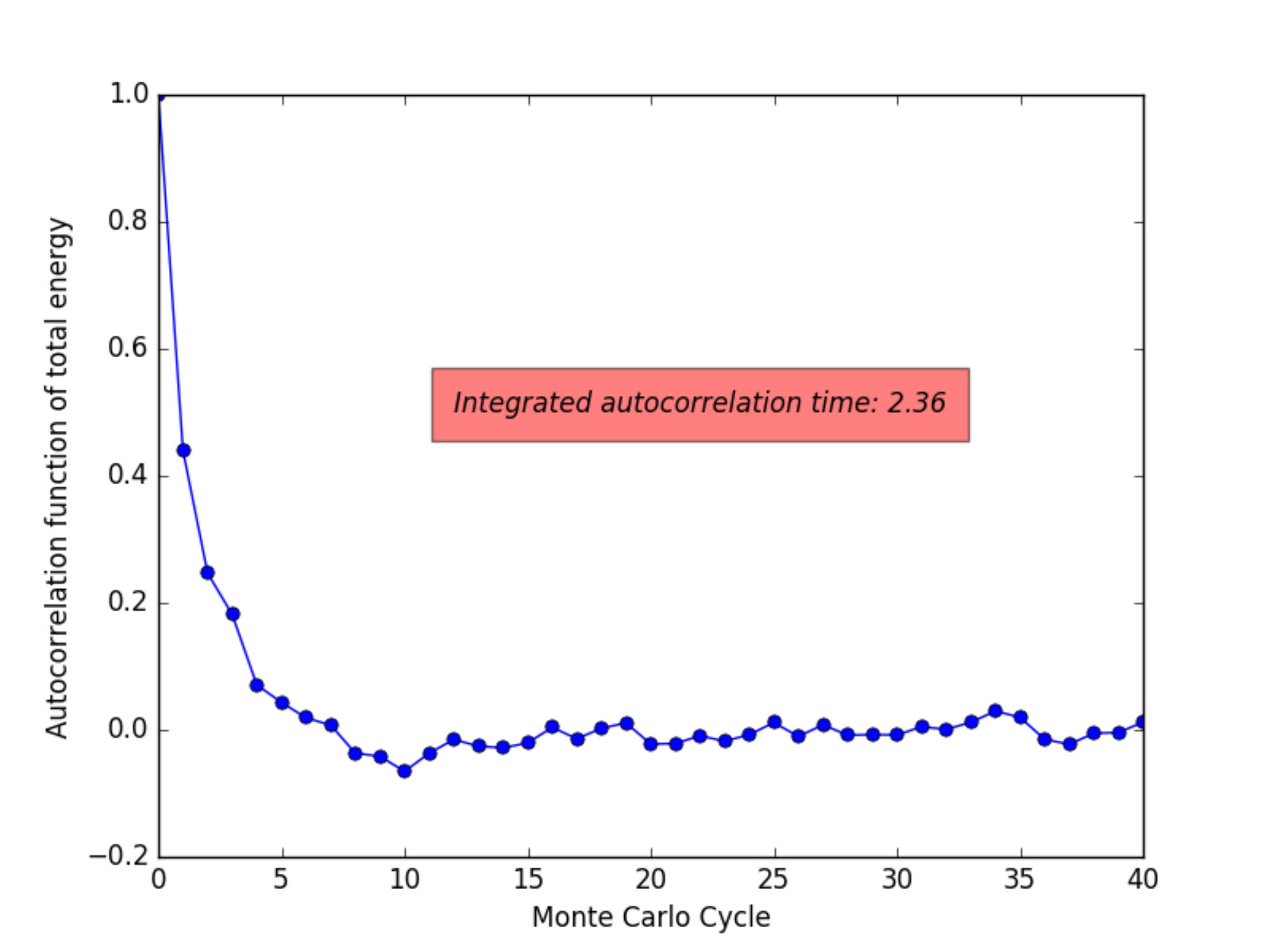}
\caption{Autocorrelation function of a 3:-1 electrolyte.}
\label{AutoC}
\end{figure}

To compute all renormalized parameters of dense electrolytes, we perform large scale simulations of electrolytes and measure all pair correlation functions.  
All our simulations are carried out in the Center for High Performance Computing (HPC) of Shanghai Jiaotong University.  To fully take advantage of computation resources, we perform one individual simulation on each GPU card.  The number of cards employed in simulation varies according to the ion densities. Typically one data point needs $2$ GPU cards, each of which carries out about $2,000$ iterations with about $5$ days.
In particular, as the system approaches the charge oscillation regime (where ion densities are high), more Monte Carlo cycles are needed and thus more cards are used.  
For each simulation(MPI Process), the memory cost on host is $263$MB, and the cost of global memory on GPU is about $32$MB.
The system contains $1,024\times1,024=1,048,576$ particles with room temperature $T = 300K$, and relative dielectric constant of the solvent is chosen to be $\epsilon = 78.3$ and the Bjerrum length $ b = 7.117\AA$. We use a spherical simulation domain with hard wall boundary conditions.   To eliminate influences from the boundary, ions that are less than ten Debye lengths away from the boundary are {\it not} used for data collection.   
The initial state is generated by setting particles uniformly inside the simulation domain with hard core repulsions. The proposed position of the selected particle is uniformly generated inside a cubic center around its original position with size about $L/3$ from experience, where $L$ is the size of simulation domain. In the warming iterations we output the total energy to monitor whether the system equilibrates or not. We find that almost all the systems we study can equilibrate in about $10$ cycles. Therefore we are sure that for all the systems the correlation between successive configurations is weak. In Fig.~\ref{AutoC} we present a typical autocorrelation function of total energy for an equilibrated 3:-1 electrolyte with classical debye length $28.53\AA$, whose autocorrelation time is about $2.3$. Here the autocorrelation function of total energy is defined as
\be
C_n=\frac{\overline{ (E_k-\overline{E})(E_{k+n}-\overline{E}) }} {\overline{ (E_k-\overline{E})^2 } }.
\ee
Here $E_k$ denotes the total energy at step $k$ and $\overline{E}$ stands for the mean energy at equilibrium. The over-line can be performed by averaging variable over both ensemble (MPI procedures) and time (steps). The integrated autocorrelation time is defined as
\be
T_{c}=\frac{1}{2}+\sum\limits_{k=1}^{\infty} C_k.
\ee

\begin{figure}[tbph!]
\centering
\includegraphics[width=0.75\textwidth]{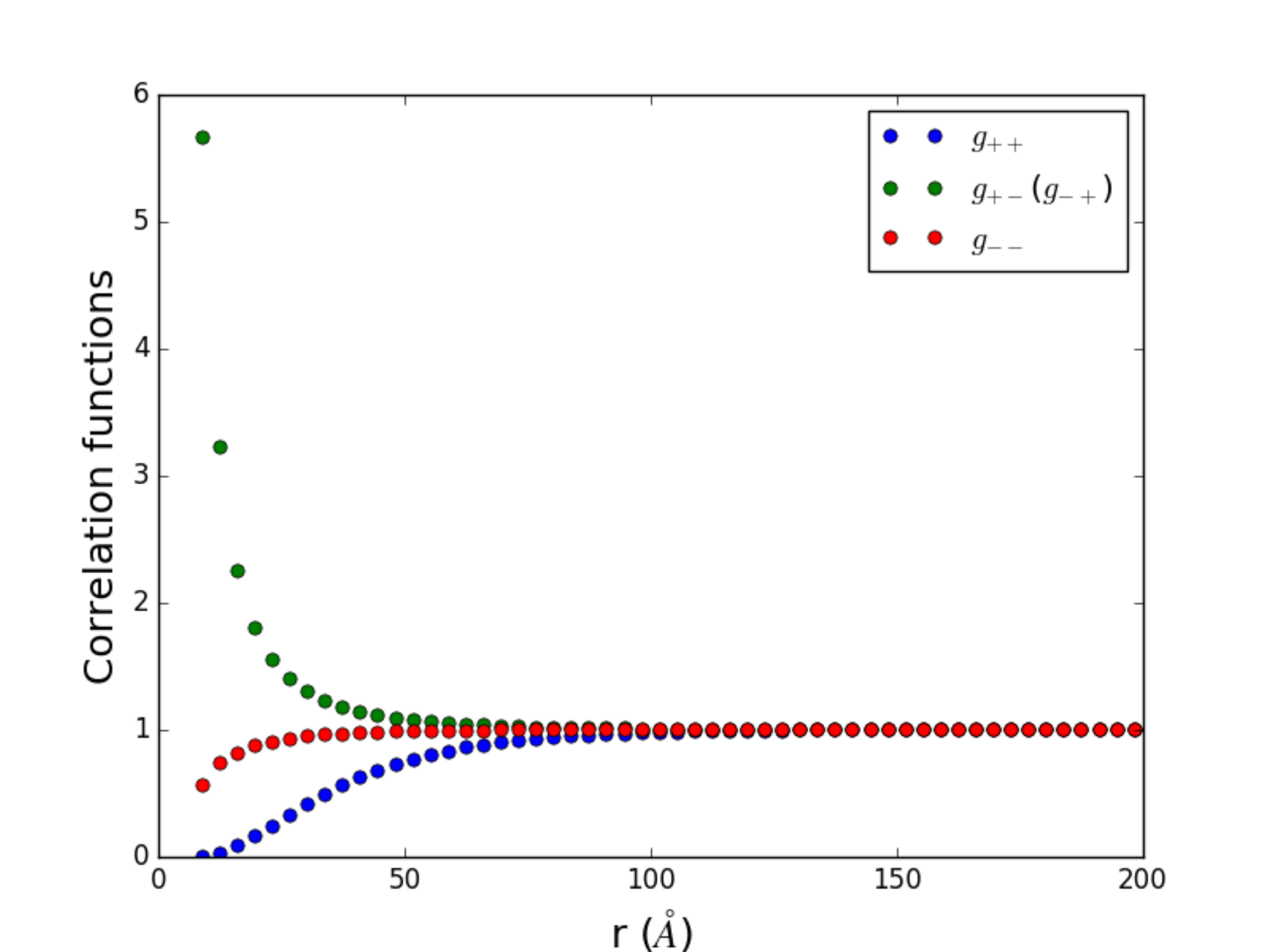}
\caption{A typical radial correlation function of 3:-1 electrolyte. The radius of ions is $3.75\AA$ and the classical debye length is $28.53\AA$.}
\label{PCF}
\end{figure}

A typical pair correlation function is showed in Fig. \ref{PCF}.

The renormalized parameters $\kappa_{\!R}, \epsilon_{\! R}, q_{\pm}^R$ can be obtained from tails of pair correlation functions $g_{\pm\pm}(\rv)$ as follows.   Firstly the two-ion PMF are obtained via:
\be
U_{\pm\pm}(\rv) = - k_B T \log g_{\pm\pm}(\rv). 
\ee
Now, let us fix a positive/negative ion $q_+/q_-$ at the origin in the bulk electrolyte.  The average charge density (excluding the  charge $q_{\pm}$ fixed at the origin) can also be computed by:
\be
\rho_{\pm}^q(\rv) = n_{+}q_+ g_{+\pm}(r)
 +n_{-} q_- g_{-\pm}(r)   . 
\label{rho-g-g}
\ee

\begin{figure}[tbph!]
\centering
\includegraphics[width=0.6\textwidth]{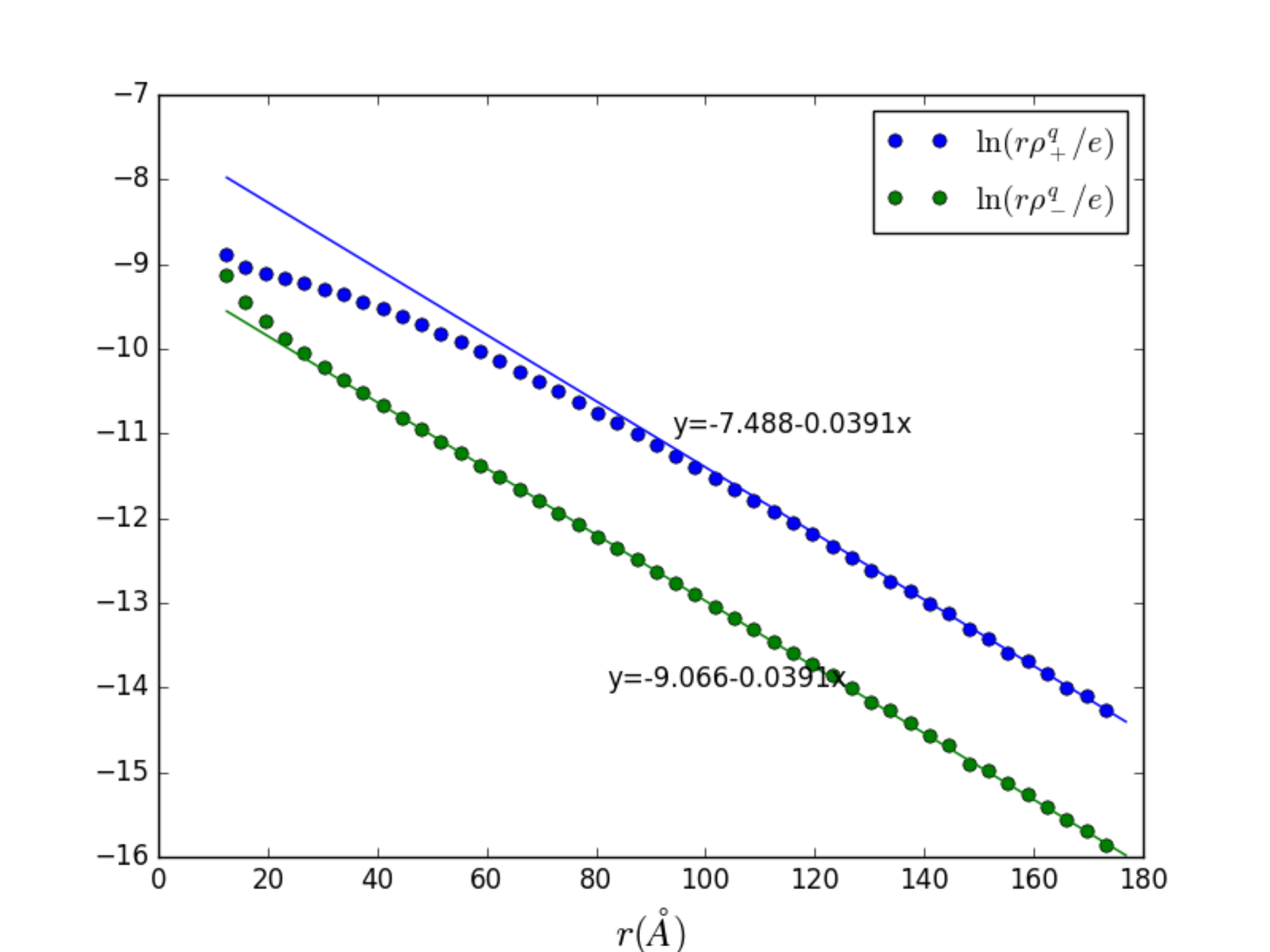}
\caption{Plotting $\ln(r\rho_\pm^q/e)$ v.s. $r$ for 3:-1 electrolyte. The radius of ions is $3.75\AA$ and the bare debye length (as prodicted by PB) is $28.53\AA$.  The renormalized Debye length is $25.57\AA$.}
\label{logrrho}
\end{figure}

Now the mean potential $\phi_{\pm}(\rv)$ around the fixed ion is given by Eq.~(\ref{phi-DIT}) with $Q_R$ replaced by $q_{\pm}^R$ in the far field.   The mean charge density $\rho_{\pm}^q(\rv) $ can be obtained using the exact Poisson equation.  In the far field, they decay in the form of screened Coulomb potential:
\ba
\rho_{\pm}^q(\rv) = - \epsilon \nabla^2 \phi_{\pm}(\rv) &\sim& 
\frac{\epsilon \kappa_{\! R}^2\, q_{\pm}^R 
e^{-\kappa_{\! R} r}}{4 \pi \epsilon_{\! R} r}. 
\label{rho_q-app} 
\ea
We can take the logarithm of Eq.~(\ref{rho_q-app}) and obtain:
\be
\log \left[ r \, \rho_{\pm}^q(\rv) \right] 
\sim \log \left[
 \frac{ \epsilon \kappa_{\! R}^2\, q_{\pm}^R }
 {4 \pi \epsilon_{\! R} }\right] 
 -\kappa_{\! R} \, r.
\label{linear-fit}
\ee 
We can therefore plot the l.h.s. (measured by simulations) as a function of radius $r$, fit the data to straight-lines {\it in the far field}, and extract the renormalized inverse Debye length $\kappa_{\! R}$ from their slopes.  This is illustrated in Fig.~(\ref{logrrho}).  Note that two straight-lines have the same slope and give a renormalized Debye length $\kappa_{\! R}^{-1} = 25.57\AA$, manifestly different from the bare Debye length $\kappa^{-1} = 28.53 \AA$. 

\begin{figure}[tbph!]
\centering
\includegraphics[width=0.7\textwidth]{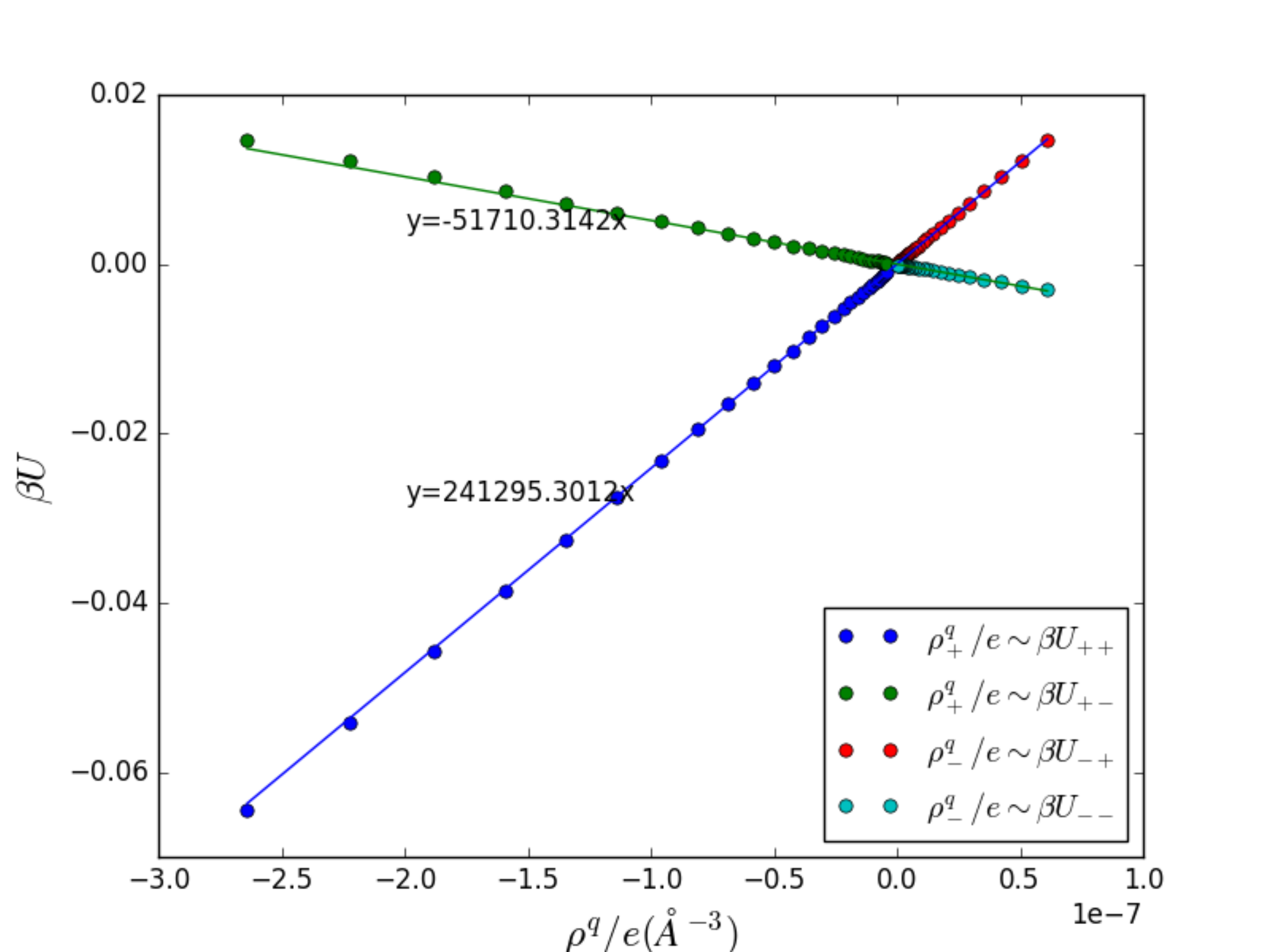}
\caption{Plotting $ \beta U$ v.s. $\rho^q/e$ for 3:-1 electrolyte. The radius of ions is $3.75\AA$ and the classical debye length is $28.53\AA$.  Using Eq.~(\ref{q_R-MC}) and fitting the long rang data to straight-lines, we obtain the renormalized valences of the positive ion and of the negative ions to be 4.125 and -0.884 respectively, which are different from the bare valences 3 and -1 respectively.  This again demonstrates the failure of classical PB theory. }
\label{rhoU}
\end{figure}

According to Eq.~(\ref{U_12-DIT}), the far field asymptotics of the two-ion PMF is
\ba
U_{\pm\pm}(\rv) &\sim& 
\frac{ q_{\pm}^R q_{\pm}^R 
e^{-\kappa_{\! R} r}}{4 \pi \epsilon_{\! R} r}. 
\label{U_pm-app} 
\ea
Taking the ratio of Eqs.~(\ref{rho_q-app}) and (\ref{U_pm-app}), we find the following relation valid in the far field:
\be
q^R_{\pm} = - \epsilon \kappa_{\! R}^2
 \frac{U_{\pm \pm}(\rv) }{\rho_{\pm}^q(\rv)}.
 \label{qR-U-rho}
\ee
Since all quantities in the r.h.s. are known, we can use this relation to determine the renormalized charges $q^R_{\pm}$ of positive and negative ions.  In fact we have two independent ways to compute the renormalized charges.  Let us write them out explicitly: 
\begin{subequations}
\label{q_R-MC}
\ba
q_+^R &=&  - \epsilon \kappa_{\! R}^2  \frac{U_{++}(\rv) }{\rho_+^q(\rv)}
=  - \epsilon \kappa_{\! R}^2  \frac{U_{-+}(\rv) }{\rho_-^q(\rv)}, \\
q_-^R &=&  - \epsilon \kappa_{\! R}^2 \frac{U_{+-}(\rv) }{\rho_+^q(\rv)}
=  - \epsilon \kappa_{\! R}^2 \frac{U_{--}(\rv) }{\rho_-^q(\rv)}. 
\ea
\end{subequations}
The data for one particular simulation are shown in Fig.~\ref{rhoU}, from which we extract the renormalized charges of positive and negative ions to be 4.125 and -0.884 respectively.  Note that these are substantially different from the bare charges, which are 3 and -1 respectively.  This charge renormalization arises as a consequence of ionic correlations and signify the failure of the classical PB theory.  

\begin{figure*}
	\centering
\subfigure[]{		\includegraphics[width=4.5cm]{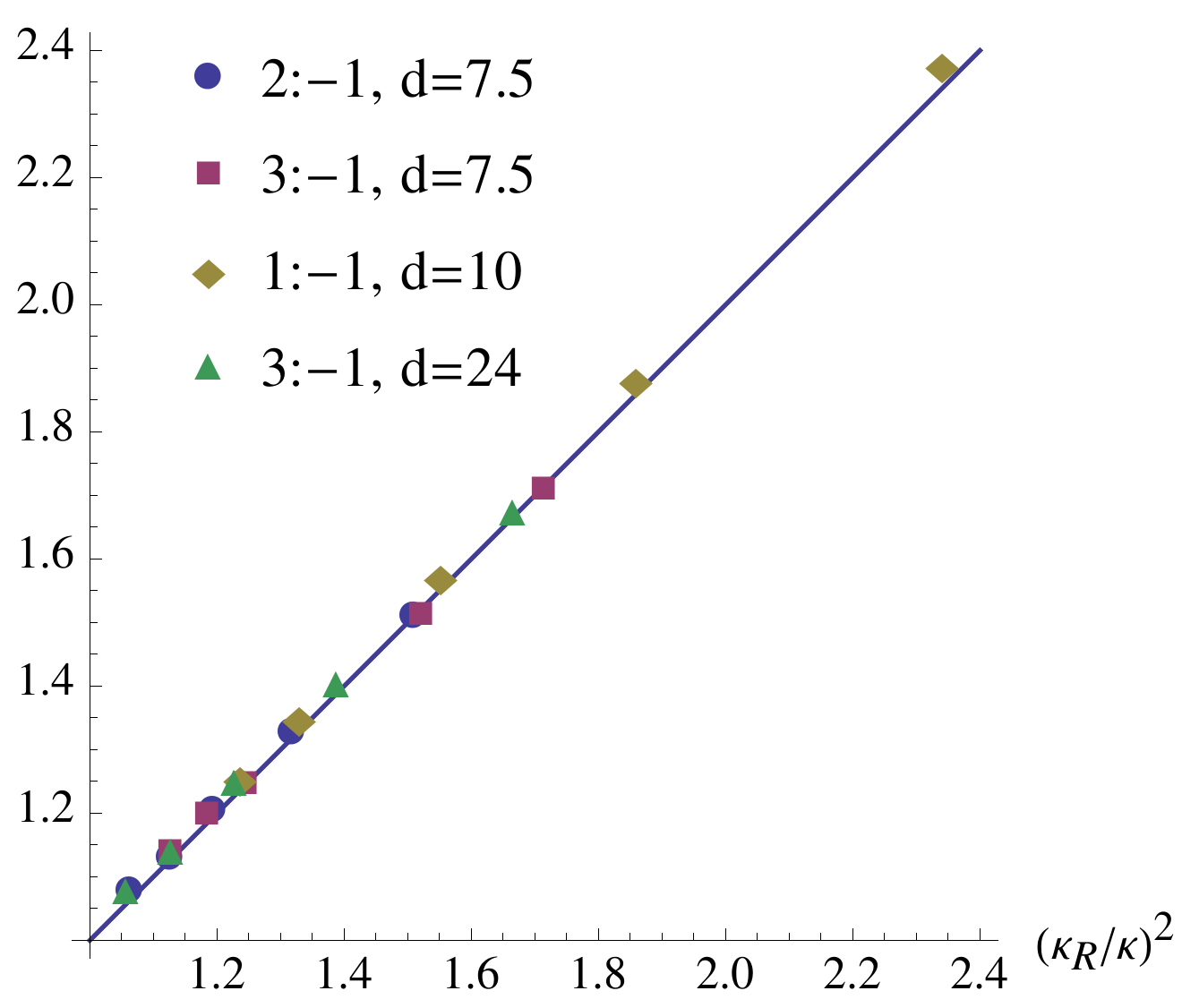}
  \label{fig:kappa_R-q_R}
}
\subfigure[]{		\includegraphics[width=4.5cm]{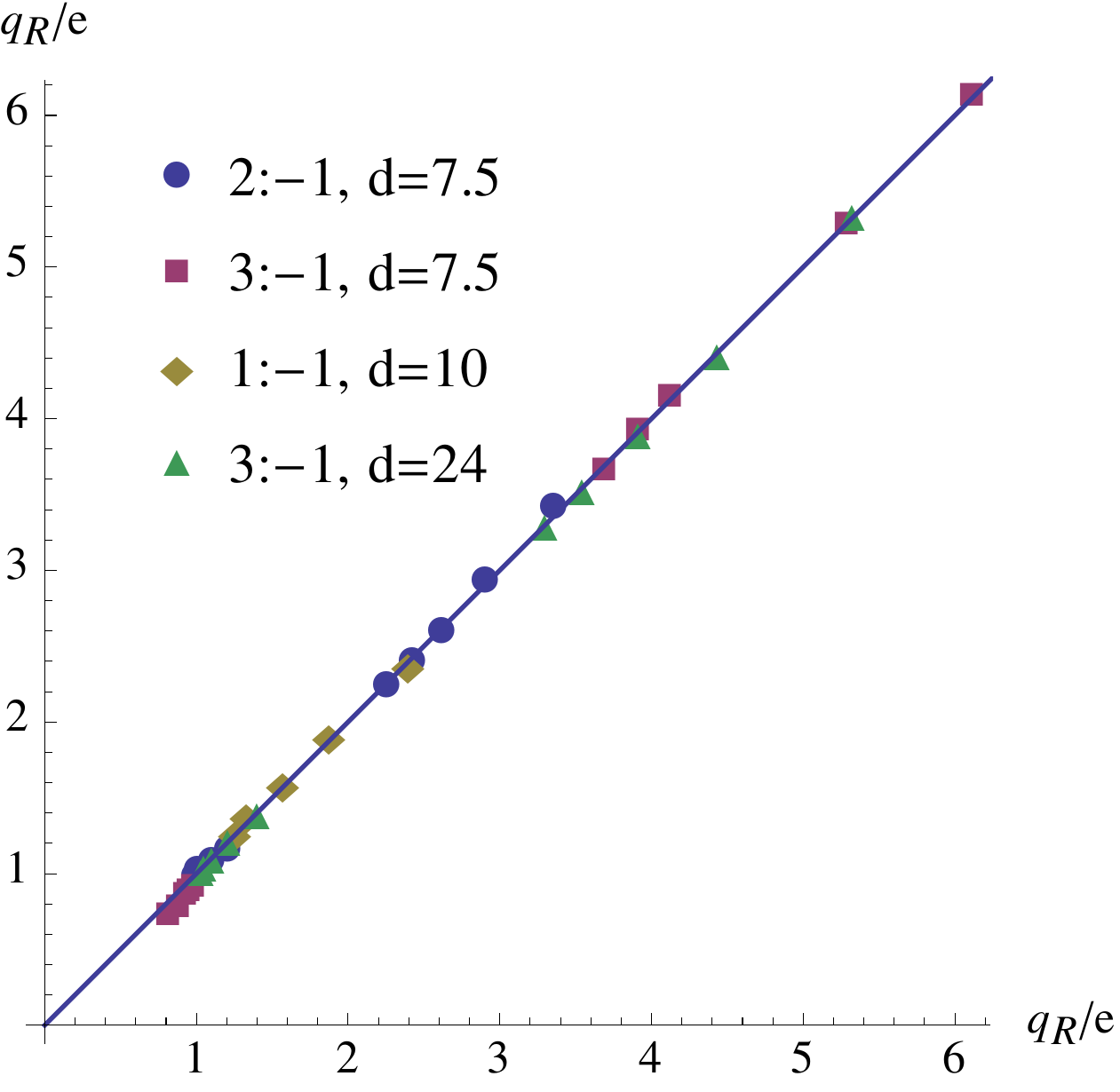}
  \label{fig:qR-qR}
}
\subfigure[]{	
				\includegraphics[width=4.5cm]{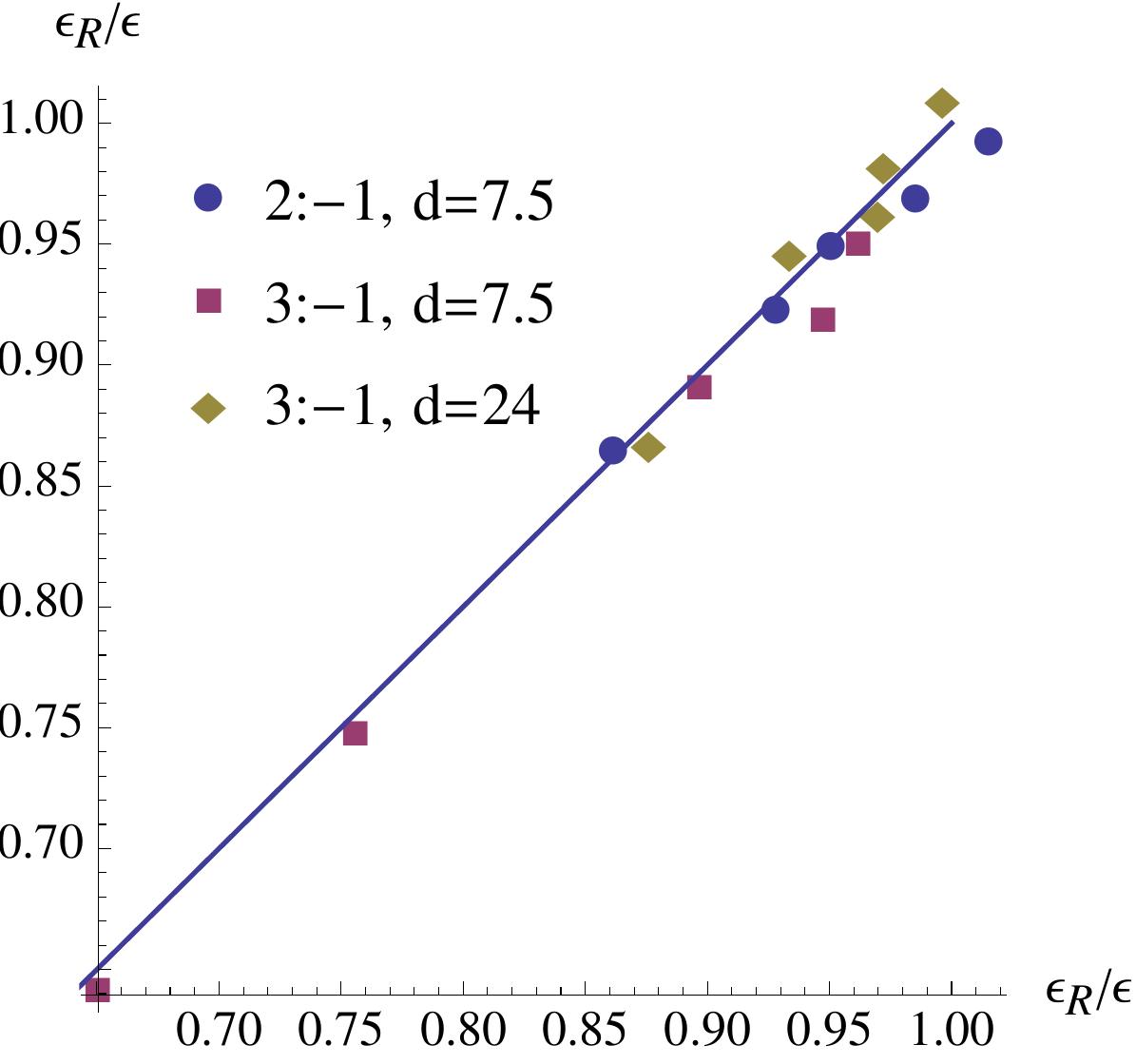}
  \label{fig:epR-epR}
}
\vspace{-3mm}
	\caption{ (a):Verification of the exact relation Eq.~(\ref{RD}) using MC simulation.  $d$ is the ion diameter (in the unit of $\AA$). Vertical axis:
	 $\left(  { q_+^{R} - q_-^{R} }\right) /  \left( { q_+ -  q_- } \right)$.  (b): The renormalized valences of ions, computed using in two independent ways, see Eqs.~(\ref{q_R-MC}).  The purpose of this panel is to demonstrate that two independent computations give the same result, within computational errors. (c): Renormalized dielectric constant can also be computed in two ways, see (\ref{ep_R-phi}).   The fact that these renormalized parameters are different from their bare values demonstrates that the properties of concentrated electrolytes are beyond the classical Poisson-Boltzmann theory. 
 \vspace{-3mm}
  }
  \label{fig:qR-epR}
\end{figure*}

Using Eq.~(\ref{rho_q-app})  we can compute the renormalized dielectric constant $\epsilon_{\! R}$ in terms of $\kappa_{\! R}, q_{\pm}^R$, and $\rho^q_{\pm}(\rv)$ (again in two independent ways):
\ba 
\epsilon_{\! R} = - \epsilon \frac{\kappa_{\!R}^2 
q_{\pm}^R e^{-\kappa_{\!R} r}}
{4 \pi  r \,\rho^q_{\pm}(\rv)}. 
\label{ep_R-phi}
\ea
Finally, we also use the computed $\kappa_{\!R}$ and $q_{\pm}^R$ to test the validity of the exact relation Eq.~(\ref{RD}).  The results are displayed in Fig.~\ref{fig:kappa_R-q_R}.  Additionally, we plot $q^R_{\pm}$ and $\kappa_{\! R}$ in Figs.~\ref{fig:qR-qR} and \ref{fig:epR-epR} respectively measured in two independent ways and show that they are consistent with each other, within computational errors.   All these numerical tests unambiguously demonstrate the validity and internal consistency of the dressed-ion theory, Eqs.~(\ref{phi-DIT})-(\ref{RD}).

\section{Conclusion}
In this work, we have developed an efficient GPU code for large-scale Monte Carlo simulation of Coulomb many-body systems, which parallelizes the sequential updating scheme and achieves an acceleration of 440 over the sequential CPU code, without sacrificing accuracy.  We have further applied this method to precisely measure the long scale linear response properties of dense asymmetric electrolytes and have demonstrated that they are beyond the classical Poisson-Boltzmann theory.  Further applications of this method will be reported in future publications.  

Y.H. Liang and X.J. Xing acknowledge financial support from NSFC (grant No. 11174196 and 91130012). Y.H. Li acknowledges support from NSF (grant No. 1066471).  The authors also thank Beijing Computational Science Research Center (BCSRC) for hospitality, where part of this work is done. This work is supported  by Center  for  HPC,  Shanghai  Jiao Tong  University.

\end{document}